%% file: Main.tex
\documentclass[nocopyrightspace]{sigplanconf}

\usepackage{color}
\usepackage{graphicx}
\usepackage{latexsym}
\usepackage{stmaryrd}
\usepackage{amsmath}
\usepackage{amssymb}
\usepackage{amsthm}
\usepackage{stmaryrd}
\usepackage{mathpartir}
\usepackage{wasysym}
\usepackage{url}
\usepackage{algorithm, algpseudocode}
\usepackage{fancyvrb}

\input{macro-comments}
\input{macros}
\input{Preamble}

\input{lemmas}

\newenvironment{example}{\par\noindent\textit{Example}\quad}{}

\begin{document}


\title{Extended Report: The Implicit Calculus}
\subtitle{A New Foundation for Generic Programming}

\authorinfo
{Bruno C. d. S. Oliveira}
{Seoul National University}
{bruno@ropas.snu.ac.kr}
\authorinfo
{Tom Schrijvers}
{Universiteit Gent}
{tom.schrijvers@ugent.be}
\authorinfo
{Wontae Choi}
{Seoul National University}
{wtchoi@ropas.snu.ac.kr}
\authorinfo
{Wonchan Lee}
{Seoul National University}
{wclee@ropas.snu.ac.kr}
\authorinfo
{Kwangkeun Yi}
{Seoul National University}
{kwang@ropas.snu.ac.kr}

\maketitle

\begin{abstract}

\input{src/Abstract}

\end{abstract}




\input{src/Introduction}
\input{src/Overview}

\input{src/Types}

\input{src/Translation}

\input{src/SourceLang}

\input{src/Related}

\input{src/Conclusion}

\paragraph{Acknowledgements}
We are grateful to Ben Delaware, Derek Dreyer, Jeremy Gibbons, Scott Kilpatrick,
Phil Wadler, Beta Ziliani, the member of ROPAS and the anonymous
reviewers for their comments and suggestions.
This work was partially supported by the Engineering Research
Center of Excellence Program of Korea Ministry of Education, Science and
Technology (MEST)~/ Korea Science and Engineering Foundation (KOSEF) grant
R11-2008-007-01002-0 and the Mid-career Researcher Program (2010-0022061)
through NRF grant funded by the MEST. This work was also partially
supported by Singapore Ministry of Education research grant 
MOE2010-T2-2-073.

\bibliography{papers}

\newpage
\appendix
\input{src/Termination}
\input{src/Proofs}

\end{document}

%% file: macro-comments.tex
\newcommand{\yank}[1]{\textbf{yank:}}



%% file: macros.tex
\newcommand{\arrow}{\rightarrow}

\newtheorem{theorem}{Theorem}[section]
\newtheorem{conjecture}{Conjecture}[section]
\newtheorem{lemma}{Lemma}[section]
\newtheorem{definition}{Definition}[section]

\newcommand{\mce}{\mathcal{E}}

\newcommand{\mcu}{\mathcal{U}}

\newcommand{\ourlang}{\lambda_{\Rightarrow}}

\newcommand{\figtwocol}[3]
{\begin{figure} #3 \caption{#2}\label{#1}\end{figure}}

\newcommand{\myirule}[2]{{\renewcommand{\arraystretch}{1.2}\ba{c} #1
                      \\ \hline #2 \ea}}

\newcommand{\ba}{\begin{array}}
\newcommand{\ea}{\end{array}}
\newcommand{\bda}{\[\ba}
\newcommand{\eda}{\ea\]}

\newcommand{\myset}[1]{\{#1\}}
\newcommand{\relation}[2]{#1:#2}

\newcommand{\To}{\Rightarrow}

\newcommand{\Abs}[2]{\Lambda #1. #2}
\newcommand{\abs}[2]{\lambda #1. #2}
\newcommand{\ruleabs}[2]{(\!|  #2  :  #1 |\!)}
\newcommand{\ruleapp}[2]{#1 {\bf ~with~} #2}
\newcommand{\type}{\tau}
\newcommand{\tyint}{{\it Int}}
\newcommand{\tybool}{{\it Bool}}
\newcommand{\tychar}{{\it Char}}

\newcommand{\tyunit}{()}
\newcommand{\unit}{()}

\newcommand{\Eq}{{\tt Eq}}

\newcommand{\rulet}{\rho}
\newcommand{\ruleschr}[3]
{\forall #1. #2 \To #3}
\newcommand{\rulesch}[3]
{\forall \vec{#1}. #2 \To #3}
\newcommand{\ruleset}{\bar{\rulet}}
\newcommand{\rulesetvar}{\ruleset}
\newcommand{\nil}{\varnothing}
\newcommand{\meta}[1]{{\it #1 }}

\newcommand{\rulepgm}{\overline{\relation{\rulet}{v}}}
\newcommand{\rulepgmvar}{\eta}

\newcommand{\rulesetexp}{\overline{\relation{e}{\rulet}}}

\newcommand{\tstate}{\mce}
\newcommand{\rclos}[1]{(#1)}
\newcommand{\lookup}[2]{{#1\langle{#2}\rangle}}

\newcommand{\denv}{\Delta}
\newcommand{\env}{\Delta}
\newcommand{\tenv}{\Gamma}
\newcommand{\eval}{\Downarrow}
\newcommand{\turns}{\vdash}
\newcommand{\vturns}{\vdash_{r}}

\newcommand{\facerule}[1]{{\tt #1}}
\newcommand{\tlabel}[1]{\mbox{(#1)}}

\newcommand{\OpQuery}{\tlabel{\facerule{OpQuery}}}
\newcommand{\OpRule}{\tlabel{\facerule{OpRule}}}
\newcommand{\OpInst}{\tlabel{\facerule{OpInst}}}
\newcommand{\OpRApp}{\tlabel{\facerule{OpRApp}}}

\newcommand{\TyQuery}{\tlabel{\facerule{TyQuery}}}
\newcommand{\TyRule}{\tlabel{\facerule{TyRule}}}
\newcommand{\TyInst}{\tlabel{\facerule{TyInst}}}
\newcommand{\TyRApp}{\tlabel{\facerule{TyRApp}}}

\newcommand{\TrInt}{\tlabel{\facerule{TrInt}}}
\newcommand{\TrVar}{\tlabel{\facerule{TrVar}}}
\newcommand{\TrAbs}{\tlabel{\facerule{TrAbs}}}
\newcommand{\TrApp}{\tlabel{\facerule{TrApp}}}
\newcommand{\TrQuery}{\tlabel{\facerule{TrQuery}}}
\newcommand{\TrRule}{\tlabel{\facerule{TrRule}}}
\newcommand{\TrInst}{\tlabel{\facerule{TrInst}}}
\newcommand{\TrRApp}{\tlabel{\facerule{TrRApp}}}

\newcommand{\SimpleRes}{\tlabel{\facerule{SimpleRes}}}
\newcommand{\RuleRes}{\tlabel{\facerule{RuleRes}}}
\newcommand{\StaRes}{\tlabel{\facerule{TyRes}}}
\newcommand{\DynRes}{\tlabel{\facerule{DynRes}}}

\newcommand{\TrRes}{\tlabel{\facerule{TrRes}}}

\newcommand{\unify}[3]{\mcu_{#3}(#1, #2)}

\newcommand{\Ex}[1]{\[\begin{array}{l}#1\end{array}\]}

\newcommand{\subst}[2]{[ #1\!\mapsto\!#2 ]}
\newcommand{\substone}{\subst{\alpha}{\type}}

\newcommand{\leteq}{\overset{{\sf let}}{=}}
\newcommand{\defeq}{\overset{{\sf def}}{=}}

\newcommand{\unambiguous}{{\sf unambiguous}}

\definecolor{gray}{gray}{0.7}
\definecolor{dark-gray}{gray}{0.4}
\definecolor{light-gray}{gray}{0.8}

\newcommand{\shade}[1]{\colorbox{light-gray}{\color{black}{$#1$}}}
\newcommand{\hide}[1]{}
\def\myruleform#1{
\setlength{\fboxrule}{0.5pt}\fbox{\normalsize $#1$}
}

\newcommand{\TyInt}{\tlabel{\facerule{TyInt}}}
\newcommand{\TyVar}{\tlabel{\facerule{TyVar}}}
\newcommand{\TyIntL}{\tlabel{\facerule{TyIntL}}}
\newcommand{\TyLVar}{\tlabel{\facerule{TyLVar}}}
\newcommand{\TyIVar}{\tlabel{\facerule{TyIVar}}}
\newcommand{\TyRec}{\tlabel{\facerule{TyRec}}}
\newcommand{\TyAbs}{\tlabel{\facerule{TyAbs}}}

\newcommand{\TyApp}{\tlabel{\facerule{TyApp}}}

\newcommand{\TyLet}{\tlabel{\facerule{TyLet}}}
\newcommand{\TyImp}{\tlabel{\facerule{TyImp}}}


%% file: lemmas.tex
\newcommand{\thmtranstypreserve}{
  \begin{theorem}[Type-preserving translation]\label{thm:type:preservation} Let $e$ be a $\ourlang$
    expression, $\type$ be a type and $E$ be a System F expression. If
    $\cdot\mid\cdot \turns \relation{e}{\type} \leadsto E$, then $\cdot \turns
    \relation{E}{|\type|}$.
  \end{theorem}
}

\newcommand{\occ}[2]{\mathit{occ}_{#1}(#2)}
\newcommand{\tnorm}[1]{|#1|}

\newcommand{\defterm}{
  \begin{definition}[Termination Condition]
    An implicit environment $\env$ satisfies the condition, denoted $\mathit{term}(\env)$, iff $\mathit{term}(\rho)$ for every $\rho = (\rulesch{\alpha}{\bar{\rho}}{\tau}) \in \mathit{dom}(\env)$, where:
    \begin{eqnarray*}
      \mathit{term}(\rho) 
      & \stackrel{\mathit{def}}{\Leftrightarrow} 
      & \occ{\alpha}{\tau'}\leq\occ{\alpha}{\tau} \\
      & & ~~~(\forall (\rulesch{\alpha'}{\bar{\rho}'}{\tau'}) \in \bar{\rho},\forall \alpha \in \mathit{ftv}(\tau,\tau_i)) \setminus \vec{\alpha}') \\
      & \wedge & \tnorm{\tau_i} < \tnorm{\tau}~~~(\forall \tau_i \in \bar{\rho}) \\
      & \wedge & \mathit{term}(\rho')~~~(\forall \rho' \in \bar{\rho})
    \end{eqnarray*}
    where
    \begin{eqnarray*}
      \occ{\alpha}{\tyint} & = & 0 \\
      \occ{\alpha}{\alpha} & = & 1 \\
      \occ{\alpha}{\alpha'} & = & 0\hspace{3cm}(\alpha \neq \alpha') \\
      \occ{\alpha}{\type_1 \rightarrow \type_2} & = & \occ{\alpha}{\type_1} + \occ{\alpha}{\type_2} \\
      \occ{\alpha}{\rulesch{\alpha}{\bar{\rho}}{\tau}} & = &  \occ{\alpha}{\tau} + \sum_{\rho\in\bar{\rho}}\occ{\alpha}{\rho} \\
      \tnorm{\tyint} & = & 1 \\
      \tnorm{\alpha} & = & 1 \\
      \tnorm{\type_1 \rightarrow \type_2} & = & 1 + \tnorm{\type_1} + \tnorm{\type_2} \\
      \tnorm{\rulesch{\alpha}{\bar{\rho}}{\tau}} & = &  1 + \tnorm{\tau} + \sum_{\rho\in\bar{\rho}}\tnorm{\rho}.
    \end{eqnarray*} 
  \end{definition}
}


%% file: src/Abstract.tex
\emph{Generic programming} (GP) is an increasingly important trend in
programming languages. Well-known GP mechanisms, such as type classes
and the C++0x concepts proposal, usually combine two features: 1) a special type of
interfaces; and 2) \emph{implicit instantiation} of implementations of
those interfaces.

Scala \emph{implicits} are a GP language mechanism, inspired by type
classes, that break with the tradition of coupling
implicit instantiation with a special type of interface. Instead,
implicits provide only implicit instantiation, which is generalized to
work for \emph{any types}. 
This turns out to be quite
powerful and useful to address many limitations that show up in other
GP mechanisms.

This paper synthesizes the key ideas of implicits formally in a minimal
and general core calculus called the implicit calculus ($\ourlang$),
and it shows how to build source languages supporting implicit
instantiation on top of it. A novelty of the calculus is its support
for \emph{partial resolution} and \emph{higher-order rules} (a feature 
that has been proposed before, but was never formalized or implemented).
Ultimately, the implicit calculus provides a formal model of implicits, 
which can be used by language designers to 
study and inform implementations of similar mechanisms in their own languages.

%% file: src/Introduction.tex
%
%
\makeatletter
\@ifundefined{lhs2tex.lhs2tex.sty.read}%
  {\@namedef{lhs2tex.lhs2tex.sty.read}{}%
   \newcommand\SkipToFmtEnd{}%
   \newcommand\EndFmtInput{}%
   \long\def\SkipToFmtEnd#1\EndFmtInput{}%
  }\SkipToFmtEnd

\newcommand\ReadOnlyOnce[1]{\@ifundefined{#1}{\@namedef{#1}{}}\SkipToFmtEnd}
\usepackage{amstext}
\usepackage{amssymb}
\usepackage{stmaryrd}
\DeclareFontFamily{OT1}{cmtex}{}
\DeclareFontShape{OT1}{cmtex}{m}{n}
  {<5><6><7><8>cmtex8
   <9>cmtex9
   <10><10.95><12><14.4><17.28><20.74><24.88>cmtex10}{}
\DeclareFontShape{OT1}{cmtex}{m}{it}
  {<-> ssub * cmtt/m/it}{}
\newcommand{\texfamily}{\fontfamily{cmtex}\selectfont}
\DeclareFontShape{OT1}{cmtt}{bx}{n}
  {<5><6><7><8>cmtt8
   <9>cmbtt9
   <10><10.95><12><14.4><17.28><20.74><24.88>cmbtt10}{}
\DeclareFontShape{OT1}{cmtex}{bx}{n}
  {<-> ssub * cmtt/bx/n}{}
\newcommand{\tex}[1]{\text{\texfamily#1}}	

\newcommand{\Sp}{\hskip.33334em\relax}

\newcommand{\Conid}[1]{\mathit{#1}}
\newcommand{\Varid}[1]{\mathit{#1}}
\newcommand{\anonymous}{\kern0.06em \vbox{\hrule\@width.5em}}
\newcommand{\plus}{\mathbin{+\!\!\!+}}
\newcommand{\bind}{\mathbin{>\!\!\!>\mkern-6.7mu=}}
\newcommand{\rbind}{\mathbin{=\mkern-6.7mu<\!\!\!<}}
\newcommand{\sequ}{\mathbin{>\!\!\!>}}
\renewcommand{\leq}{\leqslant}
\renewcommand{\geq}{\geqslant}
\usepackage{polytable}

\@ifundefined{mathindent}%
  {\newdimen\mathindent\mathindent\leftmargini}%
  {}%

\def\resethooks{%
  \global\let\SaveRestoreHook\empty
  \global\let\ColumnHook\empty}
\newcommand*{\savecolumns}[1][default]%
  {\g@addto@macro\SaveRestoreHook{\savecolumns[#1]}}
\newcommand*{\restorecolumns}[1][default]%
  {\g@addto@macro\SaveRestoreHook{\restorecolumns[#1]}}
\newcommand*{\aligncolumn}[2]%
  {\g@addto@macro\ColumnHook{\column{#1}{#2}}}

\resethooks

\newcommand{\onelinecommentchars}{\quad-{}- }
\newcommand{\commentbeginchars}{\enskip\{-}
\newcommand{\commentendchars}{-\}\enskip}

\newcommand{\visiblecomments}{%
  \let\onelinecomment=\onelinecommentchars
  \let\commentbegin=\commentbeginchars
  \let\commentend=\commentendchars}

\newcommand{\invisiblecomments}{%
  \let\onelinecomment=\empty
  \let\commentbegin=\empty
  \let\commentend=\empty}

\visiblecomments

\newlength{\blanklineskip}
\setlength{\blanklineskip}{0.66084ex}

\newcommand{\hsindent}[1]{\quad}
\let\hspre\empty
\let\hspost\empty
\newcommand{\NB}{\textbf{NB}}
\newcommand{\Todo}[1]{$\langle$\textbf{To do:}~#1$\rangle$}

\EndFmtInput
\makeatother
%
%
%
%
%
%
\ReadOnlyOnce{forall.fmt}%
\makeatletter


\let\HaskellResetHook\empty
\newcommand*{\AtHaskellReset}[1]{%
  \g@addto@macro\HaskellResetHook{#1}}
\newcommand*{\HaskellReset}{\HaskellResetHook}

\global\let\hsforallread\empty

\newcommand\hsforall{\global\let\hsdot=\hsperiodonce}
\newcommand*\hsperiodonce[2]{#2\global\let\hsdot=\hscompose}
\newcommand*\hscompose[2]{#1}

\AtHaskellReset{\global\let\hsdot=\hscompose}

\HaskellReset

\makeatother
\EndFmtInput
%
%
%
%
%
%
%
%
\ReadOnlyOnce{polycode.fmt}%
\makeatletter

\newcommand{\hsnewpar}[1]%
  {{\parskip=0pt\parindent=0pt\par\vskip #1\noindent}}

\newcommand{\hscodestyle}{}


\newcommand{\sethscode}[1]%
  {\expandafter\let\expandafter\hscode\csname #1\endcsname
   \expandafter\let\expandafter\endhscode\csname end#1\endcsname}


\newenvironment{compathscode}%
  {\par\noindent
   \advance\leftskip\mathindent
   \hscodestyle
   \let\\=\@normalcr
   \let\hspre\(\let\hspost\)%
   \pboxed}%
  {\endpboxed\)%
   \par\noindent
   \ignorespacesafterend}

\newcommand{\compaths}{\sethscode{compathscode}}


\newenvironment{plainhscode}%
  {\hsnewpar\abovedisplayskip
   \advance\leftskip\mathindent
   \hscodestyle
   \let\hspre\(\let\hspost\)%
   \pboxed}%
  {\endpboxed%
   \hsnewpar\belowdisplayskip
   \ignorespacesafterend}

\newenvironment{oldplainhscode}%
  {\hsnewpar\abovedisplayskip
   \advance\leftskip\mathindent
   \hscodestyle
   \let\\=\@normalcr
   \(\pboxed}%
  {\endpboxed\)%
   \hsnewpar\belowdisplayskip
   \ignorespacesafterend}


\newcommand{\plainhs}{\sethscode{plainhscode}}
\newcommand{\oldplainhs}{\sethscode{oldplainhscode}}
\plainhs


\newenvironment{arrayhscode}%
  {\hsnewpar\abovedisplayskip
   \advance\leftskip\mathindent
   \hscodestyle
   \let\\=\@normalcr
   \(\parray}%
  {\endparray\)%
   \hsnewpar\belowdisplayskip
   \ignorespacesafterend}

\newcommand{\arrayhs}{\sethscode{arrayhscode}}


\newenvironment{mathhscode}%
  {\parray}{\endparray}

\newcommand{\mathhs}{\sethscode{mathhscode}}


\newenvironment{texthscode}%
  {\(\parray}{\endparray\)}

\newcommand{\texths}{\sethscode{texthscode}}


\def\codeframewidth{\arrayrulewidth}
\RequirePackage{calc}

\newenvironment{framedhscode}%
  {\parskip=\abovedisplayskip\par\noindent
   \hscodestyle
   \arrayrulewidth=\codeframewidth
   \tabular{@{}|p{\linewidth-2\arraycolsep-2\arrayrulewidth-2pt}|@{}}%
   \hline\framedhslinecorrect\\{-1.5ex}%
   \let\endoflinesave=\\
   \let\\=\@normalcr
   \(\pboxed}%
  {\endpboxed\)%
   \framedhslinecorrect\endoflinesave{.5ex}\hline
   \endtabular
   \parskip=\belowdisplayskip\par\noindent
   \ignorespacesafterend}

\newcommand{\framedhslinecorrect}[2]%
  {#1[#2]}

\newcommand{\framedhs}{\sethscode{framedhscode}}


\newenvironment{inlinehscode}%
  {\(\def\column##1##2{}%
   \let\>\undefined\let\<\undefined\let\\\undefined
   \newcommand\>[1][]{}\newcommand\<[1][]{}\newcommand\\[1][]{}%
   \def\fromto##1##2##3{##3}%
   \def\nextline{}}{\) }%

\newcommand{\inlinehs}{\sethscode{inlinehscode}}


\newenvironment{joincode}%
  {\let\orighscode=\hscode
   \let\origendhscode=\endhscode
   \def\endhscode{\def\hscode{\endgroup\def\@currenvir{hscode}\\}\begingroup}
   \orighscode\def\hscode{\endgroup\def\@currenvir{hscode}}}%
  {\origendhscode
   \global\let\hscode=\orighscode
   \global\let\endhscode=\origendhscode}%

\makeatother
\EndFmtInput
%

\newcommand{\Meta}[1]{{\it #1\/}}
\newcommand{\m}[1]{\ensuremath{#1}}

\newcommand{\qlam}[2]{\m{\lambda\{#1\}.#2}}
\newcommand{\qask}[1]{\m{\Meta{?}#1}}
\newcommand{\qapp}[2]{\m{(#1)\;\texttt{with}\;#2}}
\newcommand{\qlet}[2]{\texttt{implicit}\;#1\;\texttt{in}\;#2}
\newcommand{\qLam}[2]{\m{\Lambda#1.#2}} 
  
\newcommand{\ty}[1]{\Meta{#1}}
\newcommand{\tyInt}{\Meta{int}}
\newcommand{\tyBool}{\Meta{bool}}
\newcommand{\rulety}[2]{\m{\{#1\}\!\Rightarrow\!#2}}

\section{Introduction}
\label{sec:intro}


Generic programming (GP)~\cite{musser88genericprogramming} is a
programming style that decouples algorithms from the concrete types on
which they operate. Decoupling is achieved through
parametrization. Typical forms of parametrization include
parametrization by type (for example: \emph{parametric polymorphism},
\emph{generics} or \emph{templates}) or parametrization by algebraic
structures (such as a monoid or a group).

A central idea in generic programming is \emph{implicit instantiation}
of generic parameters. Implicit instantiation means that, when generic
algorithms are called with concrete arguments, the generic arguments (concrete types, algebraic
structures, or some other form of generic parameters)
are automatically determined by the compiler.  The
benefit is that generic algorithms become as easy to use as
specialized algorithms. To illustrate implicit instantiation
and its benefits consider a \emph{polymorphic} sorting function:

\begin{hscode}\SaveRestoreHook
\column{B}{@{}>{\hspre}l<{\hspost}@{}}%
\column{3}{@{}>{\hspre}l<{\hspost}@{}}%
\column{E}{@{}>{\hspre}l<{\hspost}@{}}%
\>[3]{}\Varid{sort}\;\![\alpha\mskip1.5mu]\mathbin{:}(\alpha\to \alpha\to \Conid{Bool})\to \Conid{List}\;\alpha\to \Conid{List}\;\alpha{}\<[E]%
\ColumnHook
\end{hscode}\resethooks
with 3 parameters: the type of the elements in
the list (\ensuremath{\alpha}); the comparison operator; and the list to be
compared. Instantiating all 3 parameters explicitly at every use of
\ensuremath{\Varid{sort}} would be quite tedious.  It is likely that, for a given type,
the sorting function is called with the same, explicitly passed,
comparison function over and over again. Moreover it is easy to infer 
the type parameter \ensuremath{\alpha}.  GP greatly simplifies such calls by making
the type argument and the comparison operator implicit.
\begin{hscode}\SaveRestoreHook
\column{B}{@{}>{\hspre}l<{\hspost}@{}}%
\column{3}{@{}>{\hspre}l<{\hspost}@{}}%
\column{E}{@{}>{\hspre}l<{\hspost}@{}}%
\>[3]{}\Varid{isort}\mathbin{:}\forall \alpha\hsforall .(\alpha\to \alpha\to \Conid{Bool})\Rightarrow \Conid{List}\;\alpha\to \Conid{List}\;\alpha{}\<[E]%
\ColumnHook
\end{hscode}\resethooks
The function \ensuremath{\Varid{isort}} declares that the comparison function 
is implicit by using $\Rightarrow$ instead of $\to$. It is used as:
\begin{hscode}\SaveRestoreHook
\column{B}{@{}>{\hspre}l<{\hspost}@{}}%
\column{3}{@{}>{\hspre}l<{\hspost}@{}}%
\column{5}{@{}>{\hspre}l<{\hspost}@{}}%
\column{E}{@{}>{\hspre}l<{\hspost}@{}}%
\>[3]{}\bf{implicit}\;~\{\Varid{cmpInt}\mathbin{:}\Conid{Int}\to \Conid{Int}\to \Conid{Bool}\mskip1.5mu\}\;\mathbf{in}{}\<[E]%
\\
\>[3]{}\hsindent{2}{}\<[5]%
\>[5]{}(\Varid{isort}\;\![\mathrm{2},\mathrm{1},\mathrm{3}\mskip1.5mu],\Varid{isort}\;\![\mathrm{5},\mathrm{9},\mathrm{3}\mskip1.5mu]){}\<[E]%
\ColumnHook
\end{hscode}\resethooks
The two calls of \ensuremath{\Varid{isort}} each take only one explicit
argument: the list to be sorted.  Both the concrete type of the
elements (\ensuremath{\Conid{Int}}) and the comparison operator (\ensuremath{\Varid{cmpInt}}) are
\emph{implicitly} instantiated.

The element type is automatically inferred from the type of the list.
More interestingly, the implicit comparison operator is automatically 
determined in a process called \emph{resolution}. Resolution is a type-directed 
process that uses a set of \emph{rules}, the \emph{implicit environment}, 
to find a value that matches the type required by the function call. 
The \ensuremath{\bf{implicit}} construct extends the implicit environment with new rules.
In other words, \ensuremath{\bf{implicit}} is a \emph{scoping} construct for rules
similar to a conventional \ensuremath{\mathbf{let}}-binding. Thus, in the subexpression
\ensuremath{(\Varid{isort}\;\![\mathrm{2},\mathrm{1},\mathrm{3}\mskip1.5mu],\Varid{isort}\;\![\mathrm{5},\mathrm{9},\mathrm{3}\mskip1.5mu])}, \ensuremath{\Varid{cmpInt}} is in the local scope and 
available for resolution.

\subsection{Existing Approaches to Generic Programming}

The two main strongholds of GP are the C++ and the functional programming (FP)
communities.  Many of the pillars of GP are based on the ideas promoted
by Musser and Stepanov~\cite{musser88genericprogramming}. These ideas
were used in C++ libraries such as the Standard Template Library~\cite{musser95stl} and Boost~\cite{boost}. In
the FP community, Haskell \textit{type classes}~\cite{adhoc} have proven to be an
excellent mechanism for GP, although their original design did
not have that purpose. As years passed the FP community
created its own forms of GP~\cite{jansson96polytypic,gibbons03patternsin,sybclass}.

Garcia et al.'s~\cite{garcia03comparative} comparative study of programming language
support for GP was an important milestone for both communities.  According to that study
many languages provide some support for GP.  However, Haskell did
particularly well, largely due to type classes. A direct
consequence of that work was to bring the two main lines of work on GP
closer together and promote cross-pollination of ideas.  Haskell
adopted \emph{associated types}~\cite{assoctypes,assoctypes2}, which was the only weak point found
in the original comparison.  For the C++ community, type classes
presented an inspiration for developing language support for \emph{concepts}~\cite{musser88genericprogramming,concepts,fg}.

Several researchers started working on various approaches to concepts
(see Siek's work~\cite{siek11concepts} for a historical overview).
Some researchers focused on integrating concepts into
C++~\cite{reis06specifying,concepts}, while others focused on
developing new languages with GP in mind.  The work on System
$F^{G}$~\cite{fg,G} is an example of the latter approach: Building on
the experience from the C++ generic programming community and some of
the ideas of type classes, Siek and Lumsdaine developed a simple core
calculus based on System F which integrates concepts and improves on
type classes in several respects. In particular, System $F^{G}$
supports \emph{scoping} of rules\footnote{In the context of C++ rules
  correspond to \emph{models} or \emph{concept\_maps}.}.

During the same period Scala emerged as new contender in the area of
generic programming. Much like Haskell, Scala was not originally developed with
generic programming in mind.  However Scala included an alternative to
type classes: \emph{implicits}. Implicits were initially viewed as
\emph{a poor man's type classes}~\cite{odersky06poor}. Yet, ultimately,
they proved to be quite flexible and in some ways
superior to type classes. In fact Scala turns out to have very good
support for generic programming~\cite{scalageneric,implicits}.

A distinguishing feature of Scala implicits, and a reason for their
power, is that resolution works for
\emph{any type}. This allows Scala to simply reuse standard OO
interfaces/classes (which are regular types) to model concepts, and avoids introducing another
type of interface in the language.  In contrast, with type classes, or
the various concept proposals, resolution is tightly coupled with the type
class or concept-like interfaces.

 

\subsection{Limitations of Existing Mechanisms}

Twenty years of programming experience with type classes 
gave the FP community insights 
about the limitations of type classes. Some of these limitations 
were addressed by concept proposals. Other limitations were 
solved by implicits. However, as far as we know, no existing language 
or language proposal overcomes all limitations.
We discuss these limitations next.


\paragraph{Global scoping:} In Haskell, rules\footnote{In the context of Haskell rules correspond to \emph{type-class instances}.} are global and 
  there can be only a single rule for any given
  type~\cite{named_instance,systemct,implicit_explicit,modular}.
  Locally scoped rules are not available. Several researchers 
  have already proposed to fix this issue: with
  named rules~\cite{named_instance} or locally
  scoped ones~\cite{systemct,implicit_explicit,modular}.
  However none of those proposals have been adopted.

  Both proposals for concepts and Scala implicits offer scoping of rules
  and as such do not suffer from this limitation.

\paragraph{Second class interfaces:} Haskell type classes are second-class
  constructs compared to regular types: in Haskell, it is not possible to abstract
  over a type class~\cite{restricted}. Yet, the need for
  first-class type classes is real in practice. For example, L\"ammel and Peyton Jones~\cite{sybclass} 
  desire the following type class for their GP approach:
\begin{hscode}\SaveRestoreHook
\column{B}{@{}>{\hspre}l<{\hspost}@{}}%
\column{3}{@{}>{\hspre}l<{\hspost}@{}}%
\column{5}{@{}>{\hspre}l<{\hspost}@{}}%
\column{E}{@{}>{\hspre}l<{\hspost}@{}}%
\>[3]{}\mathbf{class}\;(\Conid{Typeable}\;\alpha,\Varid{cxt}\;\alpha)\Rightarrow \Conid{Data}\;\Varid{cxt}\;\alpha\;\mathbf{where}{}\<[E]%
\\
\>[3]{}\hsindent{2}{}\<[5]%
\>[5]{}\Varid{gmapQ}\mathbin{::}(\forall \beta\hsforall .\Conid{Data}\;\Varid{cxt}\;\beta\Rightarrow \beta\to \Varid{r})\to \alpha\to \![\Varid{r}\mskip1.5mu]{}\<[E]%
\ColumnHook
\end{hscode}\resethooks
In this type class, the intention is that the \ensuremath{\Varid{ctx}} variable abstracts over 
a concrete type class. Unfortunately, Haskell does not support type class abstraction.
Proposals for concepts inherit this limitation from type classes. 
Concepts and type classes are usually interpreted as predicates on types rather than types, 
and cannot be abstracted over as regular types. 
In contrast, because
in Scala concepts are modeled with types, 
it is possible to abstract over concepts. Oliveira and Gibbons~\cite{scalageneric} 
show how to encode this example in Scala.

\paragraph{No higher-order rules:} Finally type classes do not support 
  higher-order rules. As noted by Hinze and Peyton Jones~\cite{derivable}, 
  non-regular Haskell datatypes like:
\begin{hscode}\SaveRestoreHook
\column{B}{@{}>{\hspre}l<{\hspost}@{}}%
\column{3}{@{}>{\hspre}l<{\hspost}@{}}%
\column{E}{@{}>{\hspre}l<{\hspost}@{}}%
\>[3]{}\mathbf{data}\;\Conid{Perfect}\;\Varid{f}\;\alpha\mathrel{=}\Conid{Nil}\mid \Conid{Cons}\;\alpha\;(\Conid{Perfect}\;\Varid{f}\;(\Varid{f}\;\alpha)){}\<[E]%
\ColumnHook
\end{hscode}\resethooks
require type class instances such as:
\begin{hscode}\SaveRestoreHook
\column{B}{@{}>{\hspre}l<{\hspost}@{}}%
\column{3}{@{}>{\hspre}l<{\hspost}@{}}%
\column{5}{@{}>{\hspre}l<{\hspost}@{}}%
\column{E}{@{}>{\hspre}l<{\hspost}@{}}%
\>[3]{}\mathbf{instance}\;(\forall \beta\hsforall .\Conid{Show}\;\beta\Rightarrow \Conid{Show}\;(\Varid{f}\;\beta),\Conid{Show}\;\alpha)\Rightarrow {}\<[E]%
\\
\>[3]{}\hsindent{2}{}\<[5]%
\>[5]{}\Conid{Show}\;(\Conid{Perfect}\;\Varid{f}\;\alpha){}\<[E]%
\ColumnHook
\end{hscode}\resethooks
which Haskell does not support, as it restricts
instances (or rules) to be first-order.
This rule is \textit{higher-order} because it assumes another rule, \ensuremath{\forall \beta\hsforall .\Conid{Show}\;\beta\Rightarrow \Conid{Show}\;(\Varid{f}\;\beta)}, that contains an assumption itself. Also note that this assumed
rule is polymorphic in \ensuremath{\beta}.

  Both concept proposals and Scala implicits inherit the limitation of 
  first-order rules.

\subsection{Contributions}\label{subsec:contributions}

This paper presents $\ourlang$, a minimal and general core calculus
for implicits and it shows how to build a source language supporting
implicit instantiation on top of it. Perhaps surprisingly the core
calculus itself does not provide implicit instantiation: instantiation
of generic arguments is explicit. Instead $\ourlang$ provides two key
mechanisms for generic programming: 1) a type-directed resolution
mechanism and 2) scoping constructs for rules. Implicit instantiation
is then built as a convenience mechanism on top of $\ourlang$ by combining type-directed
resolution with conventional type-inference.  We illustrate this on a
simple, but quite expressive source language.

The calculus is inspired by Scala implicits and it
synthesizes core ideas of that mechanism formally. In
particular, like Scala implicits, a key idea is that resolution and
implicit instantiation work for any type. This allows those mechanisms 
to be more widely useful and applicable, since they can be used with
other types in the language.
The calculus is also closely related to System $F^G$, and
like System $F^G$, rules available in the
implicit environment are lexically scoped and scopes can be nested.

A novelty of our calculus is its support for partial resolution and higher-order rules. 
Although Hinze and Peyton Jones~\cite{derivable} have discussed higher-order rules informally 
and several other researchers noted their usefulness~\cite{trifonov03simulating,rodriguez08comparing,scalageneric}, no existing 
language or calculus provides support for them. Higher-order rules 
are just the analogue of higher-order functions in the implicits 
world. They arise naturally once we take the view that resolution 
should work for any type. Partial resolution adds additional expressive 
power and it is especially useful in the presence of higher-order rules.

From the GP perspective $\ourlang$
offers a new foundation for generic programming. 
The
relation between the implicit calculus and Scala implicits is
comparable to the relation between System $F^G$ and various
concept proposals; or the relation
between formal calculi of type classes and Haskell type classes: The 
implicit calculus is a minimal and general model of implicits useful for language
designers wishing to study and inform implementations of similar GP mechanisms in
their own languages.

In summary, our contributions are as follows.

\begin{itemize}
\item Our \emph{implicit calculus} $\ourlang$ 
  provides a simple, expressive and general formal model for implicits. 
  Despite its expressiveness, the
  calculus is minimal and provides an ideal setting for the formal
  study of implicits and GP.

\item Of particular interest is our resolution mechanism, which is 
  significantly more expressive than existing mechanisms 
  in the literature. It is based on a simple (logic-programming style) query language, 
  works for any type, and it supports partial resolution as well as higher-order rules.

\item The calculus has a polymorphic type system and an elaboration 
  semantics to System F. This also provides an effective implementation 
  of our calculus. The elaboration semantics is proved to be type-preserving, 
  ensuring the soundness of the calculus.

\item We present a small, but realistic source language,
  built on top of $\ourlang$ via a type-directed encoding. This language features implicit instantiation 
  and a simple type of interface, which can be used to model simple forms of 
  concepts. This source language also supports higher-order rules.

\item Finally, both $\ourlang$ and the source language have been
  implemented and the source code for their implementation is available
  at \url{http://ropas.snu.ac.kr/~bruno/implicit}.

\end{itemize}

\paragraph{Organization} 
Section 2 presents an informal overview of our calculus. 
Section 3 shows a polymorphic type system that
statically excludes ill-behaved programs. Section 4 shows the elaboration 
semantics of our calculus into System F and correctness results. 
Section 5 presents the source language and its encoding into $\ourlang$. 
Section 6 discusses comparisons and related work. Section 7 concludes.

%% file: src/Overview.tex
\makeatletter
\@ifundefined{lhs2tex.lhs2tex.sty.read}%
  {\@namedef{lhs2tex.lhs2tex.sty.read}{}%
   \newcommand\SkipToFmtEnd{}%
   \newcommand\EndFmtInput{}%
   \long\def\SkipToFmtEnd#1\EndFmtInput{}%
  }\SkipToFmtEnd

\newcommand\ReadOnlyOnce[1]{\@ifundefined{#1}{\@namedef{#1}{}}\SkipToFmtEnd}
\usepackage{amstext}
\usepackage{amssymb}
\usepackage{stmaryrd}
\DeclareFontFamily{OT1}{cmtex}{}
\DeclareFontShape{OT1}{cmtex}{m}{n}
  {<5><6><7><8>cmtex8
   <9>cmtex9
   <10><10.95><12><14.4><17.28><20.74><24.88>cmtex10}{}
\DeclareFontShape{OT1}{cmtex}{m}{it}
  {<-> ssub * cmtt/m/it}{}
\newcommand{\texfamily}{\fontfamily{cmtex}\selectfont}
\DeclareFontShape{OT1}{cmtt}{bx}{n}
  {<5><6><7><8>cmtt8
   <9>cmbtt9
   <10><10.95><12><14.4><17.28><20.74><24.88>cmbtt10}{}
\DeclareFontShape{OT1}{cmtex}{bx}{n}
  {<-> ssub * cmtt/bx/n}{}
\newcommand{\tex}[1]{\text{\texfamily#1}}	

\newcommand{\Sp}{\hskip.33334em\relax}

\newcommand{\Conid}[1]{\mathit{#1}}
\newcommand{\Varid}[1]{\mathit{#1}}
\newcommand{\anonymous}{\kern0.06em \vbox{\hrule\@width.5em}}
\newcommand{\plus}{\mathbin{+\!\!\!+}}
\newcommand{\bind}{\mathbin{>\!\!\!>\mkern-6.7mu=}}
\newcommand{\rbind}{\mathbin{=\mkern-6.7mu<\!\!\!<}}
\newcommand{\sequ}{\mathbin{>\!\!\!>}}
\renewcommand{\leq}{\leqslant}
\renewcommand{\geq}{\geqslant}
\usepackage{polytable}

\@ifundefined{mathindent}%
  {\newdimen\mathindent\mathindent\leftmargini}%
  {}%

\def\resethooks{%
  \global\let\SaveRestoreHook\empty
  \global\let\ColumnHook\empty}
\newcommand*{\savecolumns}[1][default]%
  {\g@addto@macro\SaveRestoreHook{\savecolumns[#1]}}
\newcommand*{\restorecolumns}[1][default]%
  {\g@addto@macro\SaveRestoreHook{\restorecolumns[#1]}}
\newcommand*{\aligncolumn}[2]%
  {\g@addto@macro\ColumnHook{\column{#1}{#2}}}

\resethooks

\newcommand{\onelinecommentchars}{\quad-{}- }
\newcommand{\commentbeginchars}{\enskip\{-}
\newcommand{\commentendchars}{-\}\enskip}

\newcommand{\visiblecomments}{%
  \let\onelinecomment=\onelinecommentchars
  \let\commentbegin=\commentbeginchars
  \let\commentend=\commentendchars}

\newcommand{\invisiblecomments}{%
  \let\onelinecomment=\empty
  \let\commentbegin=\empty
  \let\commentend=\empty}

\visiblecomments

\newlength{\blanklineskip}
\setlength{\blanklineskip}{0.66084ex}

\newcommand{\hsindent}[1]{\quad}
\let\hspre\empty
\let\hspost\empty
\newcommand{\NB}{\textbf{NB}}
\newcommand{\Todo}[1]{$\langle$\textbf{To do:}~#1$\rangle$}

\EndFmtInput
\makeatother
%
%
%
%
%
%
%
%
%
\ReadOnlyOnce{polycode.fmt}%
\makeatletter

\newcommand{\hsnewpar}[1]%
  {{\parskip=0pt\parindent=0pt\par\vskip #1\noindent}}

\newcommand{\hscodestyle}{}


\newcommand{\sethscode}[1]%
  {\expandafter\let\expandafter\hscode\csname #1\endcsname
   \expandafter\let\expandafter\endhscode\csname end#1\endcsname}


\newenvironment{compathscode}%
  {\par\noindent
   \advance\leftskip\mathindent
   \hscodestyle
   \let\\=\@normalcr
   \let\hspre\(\let\hspost\)%
   \pboxed}%
  {\endpboxed\)%
   \par\noindent
   \ignorespacesafterend}

\newcommand{\compaths}{\sethscode{compathscode}}


\newenvironment{plainhscode}%
  {\hsnewpar\abovedisplayskip
   \advance\leftskip\mathindent
   \hscodestyle
   \let\hspre\(\let\hspost\)%
   \pboxed}%
  {\endpboxed%
   \hsnewpar\belowdisplayskip
   \ignorespacesafterend}

\newenvironment{oldplainhscode}%
  {\hsnewpar\abovedisplayskip
   \advance\leftskip\mathindent
   \hscodestyle
   \let\\=\@normalcr
   \(\pboxed}%
  {\endpboxed\)%
   \hsnewpar\belowdisplayskip
   \ignorespacesafterend}


\newcommand{\plainhs}{\sethscode{plainhscode}}
\newcommand{\oldplainhs}{\sethscode{oldplainhscode}}
\plainhs


\newenvironment{arrayhscode}%
  {\hsnewpar\abovedisplayskip
   \advance\leftskip\mathindent
   \hscodestyle
   \let\\=\@normalcr
   \(\parray}%
  {\endparray\)%
   \hsnewpar\belowdisplayskip
   \ignorespacesafterend}

\newcommand{\arrayhs}{\sethscode{arrayhscode}}


\newenvironment{mathhscode}%
  {\parray}{\endparray}

\newcommand{\mathhs}{\sethscode{mathhscode}}


\newenvironment{texthscode}%
  {\(\parray}{\endparray\)}

\newcommand{\texths}{\sethscode{texthscode}}


\def\codeframewidth{\arrayrulewidth}
\RequirePackage{calc}

\newenvironment{framedhscode}%
  {\parskip=\abovedisplayskip\par\noindent
   \hscodestyle
   \arrayrulewidth=\codeframewidth
   \tabular{@{}|p{\linewidth-2\arraycolsep-2\arrayrulewidth-2pt}|@{}}%
   \hline\framedhslinecorrect\\{-1.5ex}%
   \let\endoflinesave=\\
   \let\\=\@normalcr
   \(\pboxed}%
  {\endpboxed\)%
   \framedhslinecorrect\endoflinesave{.5ex}\hline
   \endtabular
   \parskip=\belowdisplayskip\par\noindent
   \ignorespacesafterend}

\newcommand{\framedhslinecorrect}[2]%
  {#1[#2]}

\newcommand{\framedhs}{\sethscode{framedhscode}}


\newenvironment{inlinehscode}%
  {\(\def\column##1##2{}%
   \let\>\undefined\let\<\undefined\let\\\undefined
   \newcommand\>[1][]{}\newcommand\<[1][]{}\newcommand\\[1][]{}%
   \def\fromto##1##2##3{##3}%
   \def\nextline{}}{\) }%

\newcommand{\inlinehs}{\sethscode{inlinehscode}}


\newenvironment{joincode}%
  {\let\orighscode=\hscode
   \let\origendhscode=\endhscode
   \def\endhscode{\def\hscode{\endgroup\def\@currenvir{hscode}\\}\begingroup}
   \orighscode\def\hscode{\endgroup\def\@currenvir{hscode}}}%
  {\origendhscode
   \global\let\hscode=\orighscode
   \global\let\endhscode=\origendhscode}%

\makeatother
\EndFmtInput
%
%
%
%
%
%
\ReadOnlyOnce{forall.fmt}%
\makeatletter


\let\HaskellResetHook\empty
\newcommand*{\AtHaskellReset}[1]{%
  \g@addto@macro\HaskellResetHook{#1}}
\newcommand*{\HaskellReset}{\HaskellResetHook}

\global\let\hsforallread\empty

\newcommand\hsforall{\global\let\hsdot=\hsperiodonce}
\newcommand*\hsperiodonce[2]{#2\global\let\hsdot=\hscompose}
\newcommand*\hscompose[2]{#1}

\AtHaskellReset{\global\let\hsdot=\hscompose}

\HaskellReset

\makeatother
\EndFmtInput

\section{Overview of the Implicit Calculus $\ourlang$}
\label{sec:overview}

Our calculus $\ourlang$ combines standard scoping mechanisms 
(abstractions and applications) and types \`a la System F, with a
logic-programming-style query language. 
At the heart of the language is a threefold interpretation of types:
\begin{center}
  \ensuremath{\Varid{types}\cong\Varid{propositions}\cong\Varid{rules}}
\end{center}
\noindent Firstly, types have their traditional meaning of classifying
terms.  Secondly, via the Curry-Howard isomorphism, types can
also be interpreted as propositions -- in the context of GP, the type
proposition denotes the availability in the implicit environment of a
value of the corresponding type. Thirdly, a type is interpreted as a
logic-programming style rule, i.e., a Prolog rule or Horn
clause~\cite{kowalski}.
Resolution~\cite{resolution} connects rules and propositions: it is
the means to show (the evidence) that a proposition is entailed by a set of rules.

Next we present the key features of $\ourlang$ and how
these features are used for GP. For readability purposes we sometimes omit
redundant type annotations and slightly simplify the syntax. 

\paragraph{Fetching values by types:} A central construct in
$\ourlang$ is a query. Queries allow values to be fetched by type, not by name.  
  For example, in the following function call
\begin{hscode}\SaveRestoreHook
\column{B}{@{}>{\hspre}l<{\hspost}@{}}%
\column{3}{@{}>{\hspre}l<{\hspost}@{}}%
\column{E}{@{}>{\hspre}l<{\hspost}@{}}%
\>[3]{}\Varid{foo}\;?\!\Conid{Int}{}\<[E]%
\ColumnHook
\end{hscode}\resethooks
the query \ensuremath{?\!\Conid{Int}} looks up a value of type \ensuremath{\Conid{Int}} in the implicit
environment, to serve as an actual argument.



\paragraph{Constructing values with type-directed rules:} $\ourlang$ constructs values, using
programmer-defined, type-directed rules (similar to functions). A rule (or rule
abstraction) defines how to compute, from implicit arguments, a value of a
particular type. For example, here is a rule that computes an \ensuremath{\Conid{Int}\!\times\!\Conid{Bool}}
pair from implicit \ensuremath{\Conid{Int}} and \ensuremath{\Conid{Bool}} values:

\begin{hscode}\SaveRestoreHook
\column{B}{@{}>{\hspre}l<{\hspost}@{}}%
\column{3}{@{}>{\hspre}l<{\hspost}@{}}%
\column{E}{@{}>{\hspre}l<{\hspost}@{}}%
\>[3]{}\ruleabs{\{\mskip1.5mu \Conid{Int},\Conid{Bool}\mskip1.5mu\}\Rightarrow \Conid{Int}\!\times\!\Conid{Bool}}{(?\!\Conid{Int}\mathbin{+}\mathrm{1},\neg \;?\!\Conid{Bool})}{}\<[E]%
\ColumnHook
\end{hscode}\resethooks
The rule abstraction syntax resembles a type-annotated expression: the
expression \ensuremath{(?\!\Conid{Int}\mathbin{+}\mathrm{1},\neg \;?\!\Conid{Bool})}
to the left of the colon is
the \emph{rule body}, and to the right is the \emph{rule type} \ensuremath{\{\mskip1.5mu \Conid{Int},\Conid{Bool}\mskip1.5mu\}\Rightarrow \Conid{Int}\!\times\!\Conid{Bool}}. A rule abstraction abstracts over a set of implicit
values (here \ensuremath{\{\mskip1.5mu \Conid{Int},\Conid{Bool}\mskip1.5mu\}}), or, more generally, over rules to build 
values. 



Hence, when a value of type \ensuremath{\Conid{Int}\!\times\!\Conid{Bool}} is needed (expressed by the query \ensuremath{?\!(\Conid{Int}\!\times\!\Conid{Bool})}), the above rule can be used, provided that an integer and
a boolean value are available in the implicit environment. In such an
environment, the rule returns a pair of the incremented \ensuremath{\Conid{Int}} value and negated
\ensuremath{\Conid{Bool}} value.

The implicit environment is extended through rule application (analogous to
extending the environment with function applications).
Rule application is expressed as, for example:
\begin{hscode}\SaveRestoreHook
\column{B}{@{}>{\hspre}l<{\hspost}@{}}%
\column{3}{@{}>{\hspre}l<{\hspost}@{}}%
\column{6}{@{}>{\hspre}l<{\hspost}@{}}%
\column{E}{@{}>{\hspre}l<{\hspost}@{}}%
\>[3]{}\ruleabs{\{\mskip1.5mu \Conid{Int},\Conid{Bool}\mskip1.5mu\}\Rightarrow \Conid{Int}\!\times\!\Conid{Bool}}{(?\!\Conid{Int}\mathbin{+}\mathrm{1},\neg \;?\!\Conid{Bool})}\;{}\<[E]%
\\
\>[3]{}\hsindent{3}{}\<[6]%
\>[6]{}{\bf with}\;\{\mskip1.5mu \mathrm{1},\Conid{True}\mskip1.5mu\}{}\<[E]%
\ColumnHook
\end{hscode}\resethooks

With syntactic sugar similar to a \ensuremath{\mathbf{let}}-expression, a rule abstraction-application combination is denoted
more compactly as:

\begin{hscode}\SaveRestoreHook
\column{B}{@{}>{\hspre}l<{\hspost}@{}}%
\column{3}{@{}>{\hspre}l<{\hspost}@{}}%
\column{E}{@{}>{\hspre}l<{\hspost}@{}}%
\>[3]{}{\bf implicit}\;\{\mskip1.5mu \mathrm{1},\Conid{True}\mskip1.5mu\}\;\mathbf{in}\;(?\!\Conid{Int}\mathbin{+}\mathrm{1},\neg \;?\!\Conid{Bool}){}\<[E]%
\ColumnHook
\end{hscode}\resethooks
\noindent which returns \ensuremath{(\mathrm{2},\Conid{False})}. 

\paragraph{Higher-order rules:} $\ourlang$ supports higher-order
rules. For example, the rule 
\begin{hscode}\SaveRestoreHook
\column{B}{@{}>{\hspre}l<{\hspost}@{}}%
\column{3}{@{}>{\hspre}l<{\hspost}@{}}%
\column{E}{@{}>{\hspre}l<{\hspost}@{}}%
\>[3]{}\ruleabs{\{\mskip1.5mu \Conid{Int},\{\mskip1.5mu \Conid{Int}\mskip1.5mu\}\Rightarrow \Conid{Int}\!\times\!\Conid{Int}\mskip1.5mu\}\Rightarrow \Conid{Int}\!\times\!\Conid{Int}}{?\!(\Conid{Int}\!\times\!\Conid{Int})},{}\<[E]%
\ColumnHook
\end{hscode}\resethooks
when applied, will compute an integer pair given an integer and a rule to
compute an integer pair from an integer.
Hence, the following rule application returns $(3, 4)$:
\begin{hscode}\SaveRestoreHook
\column{B}{@{}>{\hspre}l<{\hspost}@{}}%
\column{3}{@{}>{\hspre}l<{\hspost}@{}}%
\column{5}{@{}>{\hspre}l<{\hspost}@{}}%
\column{E}{@{}>{\hspre}l<{\hspost}@{}}%
\>[3]{}{\bf implicit}\;\{\mskip1.5mu \mathrm{3},\ruleabs{\{\mskip1.5mu \Conid{Int}\mskip1.5mu\}\Rightarrow \Conid{Int}\!\times\!\Conid{Int}}{(?\!\Conid{Int},?\!\Conid{Int}\mathbin{+}\mathrm{1})}\mskip1.5mu\}\;\mathbf{in}{}\<[E]%
\\
\>[3]{}\hsindent{2}{}\<[5]%
\>[5]{}?\!(\Conid{Int}\!\times\!\Conid{Int}){}\<[E]%
\ColumnHook
\end{hscode}\resethooks
\paragraph{Recursive resolution:} 
Note that resolving the  query \ensuremath{?\!(\Conid{Int}\!\times\!\Conid{Int})}
involves applying multiple rules. 
The current environment does not contain
the required integer pair. It does however contain the integer $3$ and a rule 
\ensuremath{\ruleabs{\{\mskip1.5mu \Conid{Int}\mskip1.5mu\}\Rightarrow \Conid{Int}\!\times\!\Conid{Int}}{(?\!\Conid{Int},?\!\Conid{Int}\mathbin{+}\mathrm{1})}} to compute a
pair from an integer. Hence, the query is resolved with $(3,4)$, the
result of applying the pair-producing rule to $3$.

\paragraph{Polymorphic rules and queries:} $\ourlang$ allows polymorphic rules. For example, the rule 
\begin{hscode}\SaveRestoreHook
\column{B}{@{}>{\hspre}l<{\hspost}@{}}%
\column{3}{@{}>{\hspre}l<{\hspost}@{}}%
\column{E}{@{}>{\hspre}l<{\hspost}@{}}%
\>[3]{}\ruleabs{\forall \alpha\hsforall \hsdot{\circ }{.}\{\mskip1.5mu \alpha\mskip1.5mu\}\Rightarrow \alpha\!\times\!\alpha}{(?\!\alpha,?\!\alpha)}{}\<[E]%
\ColumnHook
\end{hscode}\resethooks
can be instantiated to multiple rules of monomorphic types

\begin{hscode}\SaveRestoreHook
\column{B}{@{}>{\hspre}l<{\hspost}@{}}%
\column{3}{@{}>{\hspre}l<{\hspost}@{}}%
\column{E}{@{}>{\hspre}l<{\hspost}@{}}%
\>[3]{}\{\mskip1.5mu \Conid{Int}\mskip1.5mu\}\Rightarrow \Conid{Int}\!\times\!\Conid{Int},\{\mskip1.5mu \Conid{Bool}\mskip1.5mu\}\Rightarrow \Conid{Bool}\!\times\!\Conid{Bool},\ldots{}\<[E]%
\ColumnHook
\end{hscode}\resethooks
Multiple monomorphic queries can be resolved by the same
rule. The following expression returns 
\ensuremath{((\mathrm{3},\mathrm{3}),(\Conid{True},\Conid{True}))}: 

\begin{hscode}\SaveRestoreHook
\column{B}{@{}>{\hspre}l<{\hspost}@{}}%
\column{3}{@{}>{\hspre}l<{\hspost}@{}}%
\column{5}{@{}>{\hspre}l<{\hspost}@{}}%
\column{23}{@{}>{\hspre}l<{\hspost}@{}}%
\column{E}{@{}>{\hspre}l<{\hspost}@{}}%
\>[3]{}{\bf implicit}\;\{\mskip1.5mu \mathrm{3},\Conid{True},{}\<[23]%
\>[23]{}\ruleabs{\forall \alpha\hsforall \hsdot{\circ }{.}\{\mskip1.5mu \alpha\mskip1.5mu\}\Rightarrow \alpha\!\times\!\alpha}{(?\!\alpha,?\!\alpha)}\mskip1.5mu\}\;\mathbf{in}{}\<[E]%
\\
\>[3]{}\hsindent{2}{}\<[5]%
\>[5]{}(?\!(\Conid{Int}\!\times\!\Conid{Int}),?\!(\Conid{Bool}\!\times\!\Conid{Bool})){}\<[E]%
\ColumnHook
\end{hscode}\resethooks
Polymorphic rules can also be used to resolve polymorphic queries:\begin{hscode}\SaveRestoreHook
\column{B}{@{}>{\hspre}l<{\hspost}@{}}%
\column{3}{@{}>{\hspre}l<{\hspost}@{}}%
\column{5}{@{}>{\hspre}l<{\hspost}@{}}%
\column{E}{@{}>{\hspre}l<{\hspost}@{}}%
\>[3]{}{\bf implicit}\;\{\mskip1.5mu \ruleabs{\forall \alpha\hsforall \hsdot{\circ }{.}\{\mskip1.5mu \alpha\mskip1.5mu\}\Rightarrow \alpha\!\times\!\alpha}{(?\!\alpha,?\!\alpha)}\mskip1.5mu\}\;\mathbf{in}{}\<[E]%
\\
\>[3]{}\hsindent{2}{}\<[5]%
\>[5]{}?\!(\forall \alpha\hsforall \hsdot{\circ }{.}\{\mskip1.5mu \alpha\mskip1.5mu\}\Rightarrow \alpha\!\times\!\alpha){}\<[E]%
\ColumnHook
\end{hscode}\resethooks
\paragraph{Combining higher-order and polymorphic rules:} 
The rule 
\begin{hscode}\SaveRestoreHook
\column{B}{@{}>{\hspre}l<{\hspost}@{}}%
\column{3}{@{}>{\hspre}l<{\hspost}@{}}%
\column{5}{@{}>{\hspre}l<{\hspost}@{}}%
\column{E}{@{}>{\hspre}l<{\hspost}@{}}%
\>[3]{}\ruleabs{\{\mskip1.5mu \Conid{Int},\forall \alpha\hsforall \hsdot{\circ }{.}\{\mskip1.5mu \alpha\mskip1.5mu\}\Rightarrow \alpha\!\times\!\alpha\mskip1.5mu\}\Rightarrow {}\<[E]%
\\
\>[3]{}\hsindent{2}{}\<[5]%
\>[5]{}(\Conid{Int}\!\times\!\Conid{Int})\!\times\!(\Conid{Int}\!\times\!\Conid{Int})}{(?\!((\Conid{Int}\!\times\!\Conid{Int})\!\times\!(\Conid{Int}\!\times\!\Conid{Int})))}{}\<[E]%
\ColumnHook
\end{hscode}\resethooks
prescribes how to build a pair of integer pairs, inductively from an
integer value, by consecutively applying the rule of type
\begin{hscode}\SaveRestoreHook
\column{B}{@{}>{\hspre}l<{\hspost}@{}}%
\column{3}{@{}>{\hspre}l<{\hspost}@{}}%
\column{E}{@{}>{\hspre}l<{\hspost}@{}}%
\>[3]{}\forall \alpha\hsforall \hsdot{\circ }{.}\{\mskip1.5mu \alpha\mskip1.5mu\}\Rightarrow \alpha\!\times\!\alpha{}\<[E]%
\ColumnHook
\end{hscode}\resethooks
twice: first to an integer, and again to the result (an
integer pair). For example, the following expression returns $((3,3),(3,3))$:

\begin{hscode}\SaveRestoreHook
\column{B}{@{}>{\hspre}l<{\hspost}@{}}%
\column{3}{@{}>{\hspre}l<{\hspost}@{}}%
\column{5}{@{}>{\hspre}l<{\hspost}@{}}%
\column{E}{@{}>{\hspre}l<{\hspost}@{}}%
\>[3]{}{\bf implicit}\;\{\mskip1.5mu \mathrm{3},\ruleabs{\forall \alpha\hsforall \hsdot{\circ }{.}\{\mskip1.5mu \alpha\mskip1.5mu\}\Rightarrow \alpha\!\times\!\alpha}{(?\!\alpha,?\!\alpha)}\mskip1.5mu\}\;\mathbf{in}{}\<[E]%
\\
\>[3]{}\hsindent{2}{}\<[5]%
\>[5]{}?\!((\Conid{Int}\!\times\!\Conid{Int})\!\times\!(\Conid{Int}\!\times\!\Conid{Int})){}\<[E]%
\ColumnHook
\end{hscode}\resethooks

\paragraph{Locally and lexically scoped rules:} 
Rules can be nested and resolution respects the lexical scope of rules. 
Consider the following program: 
\begin{hscode}\SaveRestoreHook
\column{B}{@{}>{\hspre}l<{\hspost}@{}}%
\column{3}{@{}>{\hspre}l<{\hspost}@{}}%
\column{5}{@{}>{\hspre}l<{\hspost}@{}}%
\column{E}{@{}>{\hspre}l<{\hspost}@{}}%
\>[3]{}{\bf implicit}\;\{\mskip1.5mu \mathrm{1}\mskip1.5mu\}\;\mathbf{in}{}\<[E]%
\\
\>[3]{}\hsindent{2}{}\<[5]%
\>[5]{}{\bf implicit}\;\{\mskip1.5mu \Conid{True},\ruleabs{\{\mskip1.5mu \Conid{Bool}\mskip1.5mu\}\Rightarrow \Conid{Int}}{\;\mathbf{if}\;?\!\Conid{Bool}\;\mathbf{then}\;\mathrm{2}}\mskip1.5mu\}{}\<[E]%
\\
\>[3]{}\hsindent{2}{}\<[5]%
\>[5]{}\mathbf{in}\;?\!\Conid{Int}{}\<[E]%
\ColumnHook
\end{hscode}\resethooks
The query $\qask{\tyint}$ is not resolved with the
integer value \ensuremath{\mathrm{1}}. Instead the rule that returns an integer from a boolean
is applied to the boolean \ensuremath{\Conid{True}}, because those two rules 
can provide an integer value and they are nearer to the query. So, 
the program returns $2$ and not $1$.

\paragraph{Overlapping rules:} 
Two rules overlap if their return types intersect, i.e., when they can both 
be used to resolve the same query. Overlapping rules are
allowed in $\ourlang$ through nested scoping. The nearest matching
rule takes priority over other matching rules. For example consider 
the following program:

\begin{hscode}\SaveRestoreHook
\column{B}{@{}>{\hspre}l<{\hspost}@{}}%
\column{3}{@{}>{\hspre}l<{\hspost}@{}}%
\column{6}{@{}>{\hspre}l<{\hspost}@{}}%
\column{9}{@{}>{\hspre}l<{\hspost}@{}}%
\column{E}{@{}>{\hspre}l<{\hspost}@{}}%
\>[3]{}{\bf implicit}\;\{\mskip1.5mu \lambda\Varid{x}.\Varid{x}\mathbin{:}\forall \alpha\hsforall \hsdot{\circ }{.}\alpha\to \alpha\mskip1.5mu\}\;\mathbf{in}{}\<[E]%
\\
\>[3]{}\hsindent{3}{}\<[6]%
\>[6]{}{\bf implicit}\;\{\mskip1.5mu \lambda\Varid{n}.\Varid{n}\mathbin{+}\mathrm{1}\mathbin{:}\Conid{Int}\to \Conid{Int}\mskip1.5mu\}\;\mathbf{in}{}\<[E]%
\\
\>[6]{}\hsindent{3}{}\<[9]%
\>[9]{}?\!(\Conid{Int}\to \Conid{Int})\;\mathrm{1}{}\<[E]%
\ColumnHook
\end{hscode}\resethooks
In this case \ensuremath{\lambda\Varid{n}.\Varid{n}\mathbin{+}\mathrm{1}\mathbin{:}\Conid{Int}\to \Conid{Int}} is the lexically nearest
match in the implicit environment and evaluating this program results 
in \ensuremath{\mathrm{2}}. However, if we have the following program instead:
\begin{hscode}\SaveRestoreHook
\column{B}{@{}>{\hspre}l<{\hspost}@{}}%
\column{3}{@{}>{\hspre}l<{\hspost}@{}}%
\column{5}{@{}>{\hspre}l<{\hspost}@{}}%
\column{8}{@{}>{\hspre}l<{\hspost}@{}}%
\column{E}{@{}>{\hspre}l<{\hspost}@{}}%
\>[3]{}{\bf implicit}\;\{\mskip1.5mu \lambda\Varid{n}.\Varid{n}\mathbin{+}\mathrm{1}\mathbin{:}\Conid{Int}\to \Conid{Int}\mskip1.5mu\}\;\mathbf{in}{}\<[E]%
\\
\>[3]{}\hsindent{2}{}\<[5]%
\>[5]{}{\bf implicit}\;\{\mskip1.5mu \lambda\Varid{x}.\Varid{x}\mathbin{:}\forall \alpha\hsforall \hsdot{\circ }{.}\alpha\to \alpha\mskip1.5mu\}\;\mathbf{in}{}\<[E]%
\\
\>[5]{}\hsindent{3}{}\<[8]%
\>[8]{}?\!(\Conid{Int}\to \Conid{Int})\;\mathrm{1}{}\<[E]%
\ColumnHook
\end{hscode}\resethooks
Then the lexically nearest match is \ensuremath{\lambda\Varid{x}.\Varid{x}\mathbin{:}\forall \alpha\hsforall \hsdot{\circ }{.}\alpha\to \alpha}
and evaluating this program results in \ensuremath{\mathrm{1}}.



%% file: src/Types.tex
%
%
\makeatletter
\@ifundefined{lhs2tex.lhs2tex.sty.read}%
  {\@namedef{lhs2tex.lhs2tex.sty.read}{}%
   \newcommand\SkipToFmtEnd{}%
   \newcommand\EndFmtInput{}%
   \long\def\SkipToFmtEnd#1\EndFmtInput{}%
  }\SkipToFmtEnd

\newcommand\ReadOnlyOnce[1]{\@ifundefined{#1}{\@namedef{#1}{}}\SkipToFmtEnd}
\usepackage{amstext}
\usepackage{amssymb}
\usepackage{stmaryrd}
\DeclareFontFamily{OT1}{cmtex}{}
\DeclareFontShape{OT1}{cmtex}{m}{n}
  {<5><6><7><8>cmtex8
   <9>cmtex9
   <10><10.95><12><14.4><17.28><20.74><24.88>cmtex10}{}
\DeclareFontShape{OT1}{cmtex}{m}{it}
  {<-> ssub * cmtt/m/it}{}
\newcommand{\texfamily}{\fontfamily{cmtex}\selectfont}
\DeclareFontShape{OT1}{cmtt}{bx}{n}
  {<5><6><7><8>cmtt8
   <9>cmbtt9
   <10><10.95><12><14.4><17.28><20.74><24.88>cmbtt10}{}
\DeclareFontShape{OT1}{cmtex}{bx}{n}
  {<-> ssub * cmtt/bx/n}{}
\newcommand{\tex}[1]{\text{\texfamily#1}}	

\newcommand{\Sp}{\hskip.33334em\relax}

\newcommand{\Conid}[1]{\mathit{#1}}
\newcommand{\Varid}[1]{\mathit{#1}}
\newcommand{\anonymous}{\kern0.06em \vbox{\hrule\@width.5em}}
\newcommand{\plus}{\mathbin{+\!\!\!+}}
\newcommand{\bind}{\mathbin{>\!\!\!>\mkern-6.7mu=}}
\newcommand{\rbind}{\mathbin{=\mkern-6.7mu<\!\!\!<}}
\newcommand{\sequ}{\mathbin{>\!\!\!>}}
\renewcommand{\leq}{\leqslant}
\renewcommand{\geq}{\geqslant}
\usepackage{polytable}

\@ifundefined{mathindent}%
  {\newdimen\mathindent\mathindent\leftmargini}%
  {}%

\def\resethooks{%
  \global\let\SaveRestoreHook\empty
  \global\let\ColumnHook\empty}
\newcommand*{\savecolumns}[1][default]%
  {\g@addto@macro\SaveRestoreHook{\savecolumns[#1]}}
\newcommand*{\restorecolumns}[1][default]%
  {\g@addto@macro\SaveRestoreHook{\restorecolumns[#1]}}
\newcommand*{\aligncolumn}[2]%
  {\g@addto@macro\ColumnHook{\column{#1}{#2}}}

\resethooks

\newcommand{\onelinecommentchars}{\quad-{}- }
\newcommand{\commentbeginchars}{\enskip\{-}
\newcommand{\commentendchars}{-\}\enskip}

\newcommand{\visiblecomments}{%
  \let\onelinecomment=\onelinecommentchars
  \let\commentbegin=\commentbeginchars
  \let\commentend=\commentendchars}

\newcommand{\invisiblecomments}{%
  \let\onelinecomment=\empty
  \let\commentbegin=\empty
  \let\commentend=\empty}

\visiblecomments

\newlength{\blanklineskip}
\setlength{\blanklineskip}{0.66084ex}

\newcommand{\hsindent}[1]{\quad}
\let\hspre\empty
\let\hspost\empty
\newcommand{\NB}{\textbf{NB}}
\newcommand{\Todo}[1]{$\langle$\textbf{To do:}~#1$\rangle$}

\EndFmtInput
\makeatother
%
%
%
%
%
%
\ReadOnlyOnce{forall.fmt}%
\makeatletter


\let\HaskellResetHook\empty
\newcommand*{\AtHaskellReset}[1]{%
  \g@addto@macro\HaskellResetHook{#1}}
\newcommand*{\HaskellReset}{\HaskellResetHook}

\global\let\hsforallread\empty

\newcommand\hsforall{\global\let\hsdot=\hsperiodonce}
\newcommand*\hsperiodonce[2]{#2\global\let\hsdot=\hscompose}
\newcommand*\hscompose[2]{#1}

\AtHaskellReset{\global\let\hsdot=\hscompose}

\HaskellReset

\makeatother
\EndFmtInput
%
%
%
%
%
%
%
%
\ReadOnlyOnce{polycode.fmt}%
\makeatletter

\newcommand{\hsnewpar}[1]%
  {{\parskip=0pt\parindent=0pt\par\vskip #1\noindent}}

\newcommand{\hscodestyle}{}


\newcommand{\sethscode}[1]%
  {\expandafter\let\expandafter\hscode\csname #1\endcsname
   \expandafter\let\expandafter\endhscode\csname end#1\endcsname}


\newenvironment{compathscode}%
  {\par\noindent
   \advance\leftskip\mathindent
   \hscodestyle
   \let\\=\@normalcr
   \let\hspre\(\let\hspost\)%
   \pboxed}%
  {\endpboxed\)%
   \par\noindent
   \ignorespacesafterend}

\newcommand{\compaths}{\sethscode{compathscode}}


\newenvironment{plainhscode}%
  {\hsnewpar\abovedisplayskip
   \advance\leftskip\mathindent
   \hscodestyle
   \let\hspre\(\let\hspost\)%
   \pboxed}%
  {\endpboxed%
   \hsnewpar\belowdisplayskip
   \ignorespacesafterend}

\newenvironment{oldplainhscode}%
  {\hsnewpar\abovedisplayskip
   \advance\leftskip\mathindent
   \hscodestyle
   \let\\=\@normalcr
   \(\pboxed}%
  {\endpboxed\)%
   \hsnewpar\belowdisplayskip
   \ignorespacesafterend}


\newcommand{\plainhs}{\sethscode{plainhscode}}
\newcommand{\oldplainhs}{\sethscode{oldplainhscode}}
\plainhs


\newenvironment{arrayhscode}%
  {\hsnewpar\abovedisplayskip
   \advance\leftskip\mathindent
   \hscodestyle
   \let\\=\@normalcr
   \(\parray}%
  {\endparray\)%
   \hsnewpar\belowdisplayskip
   \ignorespacesafterend}

\newcommand{\arrayhs}{\sethscode{arrayhscode}}


\newenvironment{mathhscode}%
  {\parray}{\endparray}

\newcommand{\mathhs}{\sethscode{mathhscode}}


\newenvironment{texthscode}%
  {\(\parray}{\endparray\)}

\newcommand{\texths}{\sethscode{texthscode}}


\def\codeframewidth{\arrayrulewidth}
\RequirePackage{calc}

\newenvironment{framedhscode}%
  {\parskip=\abovedisplayskip\par\noindent
   \hscodestyle
   \arrayrulewidth=\codeframewidth
   \tabular{@{}|p{\linewidth-2\arraycolsep-2\arrayrulewidth-2pt}|@{}}%
   \hline\framedhslinecorrect\\{-1.5ex}%
   \let\endoflinesave=\\
   \let\\=\@normalcr
   \(\pboxed}%
  {\endpboxed\)%
   \framedhslinecorrect\endoflinesave{.5ex}\hline
   \endtabular
   \parskip=\belowdisplayskip\par\noindent
   \ignorespacesafterend}

\newcommand{\framedhslinecorrect}[2]%
  {#1[#2]}

\newcommand{\framedhs}{\sethscode{framedhscode}}


\newenvironment{inlinehscode}%
  {\(\def\column##1##2{}%
   \let\>\undefined\let\<\undefined\let\\\undefined
   \newcommand\>[1][]{}\newcommand\<[1][]{}\newcommand\\[1][]{}%
   \def\fromto##1##2##3{##3}%
   \def\nextline{}}{\) }%

\newcommand{\inlinehs}{\sethscode{inlinehscode}}


\newenvironment{joincode}%
  {\let\orighscode=\hscode
   \let\origendhscode=\endhscode
   \def\endhscode{\def\hscode{\endgroup\def\@currenvir{hscode}\\}\begingroup}
   \orighscode\def\hscode{\endgroup\def\@currenvir{hscode}}}%
  {\origendhscode
   \global\let\hscode=\orighscode
   \global\let\endhscode=\origendhscode}%

\makeatother
\EndFmtInput

\newcommand{\rhs}[1]{\mathit{rhs}(#1)}
\newcommand{\lhs}[1]{\mathit{lhs}(#1)}
\newcommand{\qtv}[1]{\mathit{qtv}(#1)}
\newcommand{\ftv}[1]{\mathit{ftv}(#1)}

\section{The $\ourlang$ Calculus}
\label{sec:ourlang}

This section formalizes the syntax and type system of $\ourlang$. 

\subsection{Syntax}    
\label{subsec:syntax}

This is the syntax of the calculus:
{\bda{llrl}
    \text{(Simple) Types} & \type \hide{\in \meta{Type}} & ::=  & \alpha \mid \tyint \mid \type_1 \arrow \type_2 \mid \rulet \\
    \text{Rule Types} & \rulet \hide{\in \meta{RType}} & ::= & 
    \rulesch{\alpha}{\rulesetvar}{\type} \\
    \text{Expressions} & \ensuremath{\Varid{e}} & ::=  &
    n \mid x \mid \lambda x:\type.e \mid e_1\,e_2 \\
    & & \mid &
    \ensuremath{?\!\rulet} \mid 
    \ensuremath{\ruleabs{\rulet}{\Varid{e}}} \mid
    e[\vec{\type}] \mid
    \ensuremath{\Varid{e}\;{\bf with}\;\rulesetexp} \\
  \eda }

\textit{Types} $\type$ are either type variables $\alpha$, the integer type
$\tyint$, function types $\type_1 \arrow \type_2$ or rule types $\rulet$. 
A \textit{rule type} $\rulet = \rulesch{\alpha}{\rulesetvar}{\type}$
is a type scheme with universally quantified variables $\vec{\alpha}$
and an (implicit) \textit{context} $\rulesetvar$.
This \textit{context} summarizes the assumed implicit environment. 
Note that we use $\vec{o}$ to denote an ordered sequence $o_1,\ldots,o_n$ of
entities and $\bar{o}$ to denote a set $\{o_1,\ldots,o_n\}$.
Such ordered sequences and sets can be empty, and 
we often omit empty universal quantifiers and empty contexts from a
rule type. The base case of rule types is when 
$\rulesetvar$ is the empty set ($\rulesch{\alpha}{\{\}}{\type}$). 

Expressions include integer constants $n$ and the three basic typed
$\lambda$-calculus expressions (variables, lambda binders and applications). 
A \textit{query} \ensuremath{?\!\rulet} queries
the implicit environment for a value of type $\rulet$.  A \textit{rule
abstraction} \ensuremath{\ruleabs{\rulesch{\alpha}{\rulesetvar}{\type}}{\Varid{e}}} builds a rule whose type is \ensuremath{\rulesch{\alpha}{\rulesetvar}{\type}}
and whose body is $e$. 

Without loss of generality we assume that all variables $x$
and type variables $\alpha$ in binders are distinct. If not, they
can be easily renamed apart to be so.


Note that, unlike System F, our calculus does not have a 
separate $\Lambda$ binder for type variables. Instead 
rule abstractions play a dual role in the binding structure: 1) the universal 
quantification of type variables (which binds types), and 2) the context 
(which binds a rule set). This design choice is due to our interpretation of 
rules as logic programming rules\footnote{In Prolog these
are not separated either.}. After all, in the matching process of resolution, a
rule is applied as a unit.  Hence, separating rules into more primitive binders
(\`a la System F's type and value binders) would only complicate the definition
of resolution unnecessarily. However, elimination can be modularized into 
two constructs: \textit{type application} $e[\bar{\type}]$ and \textit{rule
application} \ensuremath{\Varid{e}\;{\bf with}\;\rulesetexp}.

Using rule abstractions and applications we can build the \ensuremath{{\bf implicit}} 
sugar that we have used in Sections~\ref{sec:intro} and \ref{sec:overview}.

\begin{hscode}\SaveRestoreHook
\column{B}{@{}>{\hspre}l<{\hspost}@{}}%
\column{3}{@{}>{\hspre}l<{\hspost}@{}}%
\column{E}{@{}>{\hspre}l<{\hspost}@{}}%
\>[3]{}{\bf implicit}\;{ \overline{e : \rho}}\;\mathbf{in}\;\Varid{e_1}\mathbin{:}\tau\defeq\ruleabs{{ \overline{\rho}}\Rightarrow \tau}{\Varid{e_1}}\;{\bf with}\;{ \overline{e : \rho}}{}\<[E]%
\ColumnHook
\end{hscode}\resethooks

For readability purposes, when we use \ensuremath{{\bf implicit}} we omit the type annotation 
\ensuremath{\tau}. As we shall see in Section~\ref{sec:example} this annotation can be 
automatically inferred.

For brevity and simplicity reasons, we have kept $\ourlang$ small.  In examples
we may use additional syntax such as built-in integer operators and boolean
literals and types. 

%
\subsection{Type System}
\label{sec:types}




\figtwocol{fig:type}{Type System}{
\small
\bda{llrl} 
\text{Type Environments} & \tenv & ::= & \cdot \mid \tenv; \relation{x}{\type} \\
\text{Implicit Environments} & \env & ::= & \cdot \mid \env; \rulesetvar \\
\eda \\
\bda{lc}
\multicolumn{2}{l}{\myruleform{\tenv\mid\env \turns \relation{e}{\type}}} \\

\TyInt &
{ \tenv\mid\env \turns \relation{n}{\tyint} } 
\\ \\

\TyVar &
\myirule
{ (x:\type) \in \tenv 
}
{ \tenv\mid\env \turns \relation{x}{\type} } 
\\ \\

\TyAbs &
\myirule
{ \tenv;\relation{x}{\type_1}\mid \env \turns \relation{e}{\type_2}
}
{ \tenv\mid\env \turns \relation{\lambda \relation{x}{\type_1}.e}{\type_1 \arrow \type_2} } 
\\ \\

\TyApp &
\myirule
{ \tenv\mid \env \turns \relation{e_1}{\type_2 \arrow \type_1} \quad\quad
  \tenv\mid \env \turns \relation{e_2}{\type_2}
}
{ \tenv\mid\env \turns \relation{e_1\,e_2}{\type_1} } 
\\ \\

\TyRule &
\myirule
{ \rulet = \rulesch{\alpha}{\rulesetvar}{\type} \quad
  \shade{\unambiguous(\rulet)} \\
  \tenv\mid \env; \rulesetvar \turns \relation{e}{\type} \quad
  \vec{\alpha} \cap \mathit{ftv}(\tenv,\env) = \emptyset
}
{ \tenv\mid \env \turns \relation{\ruleabs{\rulet}{e}}{\rulet}
} 
\\ \\

\TyInst &
\myirule
{ \tenv\mid \env \turns 
  \relation{e}{\rulesch{\alpha}{\rulesetvar}{\type}}
}
{ \tenv\mid \env \turns
  \relation{e[\vec{\type}]}
  {\subst{\vec{\alpha}}{\vec{\type}}(\rulesetvar \To \type)}
}
\\ \\

\TyRApp &
\myirule
{ \tenv\mid \env \turns \relation{e}{\ruleset \To \type} \\
  \tenv\mid \env \turns \relation{e_i}{\rulet_i} \quad
  (\forall \relation{e_i}{\rulet_i} \in
  \overline{\relation{e}{\rulet}})
}
{ \tenv\mid \env \turns 
  \relation{(\ruleapp{e}{\rulesetexp})}{\type}
} \\ \\

\TyQuery &
\myirule
{ \env \vturns \rulet 
  \quad \shade{\unambiguous(\rulet)}
}
{ \tenv\mid \env \turns \relation{?\rulet}{\rulet} }

\\ \\

\multicolumn{2}{l}{\myruleform{\env \vturns \rulet}} \\

\StaRes &
\myirule
{ \lookup{\env}{\type} = \rulesetvar' \Rightarrow \type \\
  \env \vturns \rulet_i \quad 
  (\forall \rulet_i \in \rulesetvar' - \rulesetvar)
}
{ \env \vturns \rulesch{\alpha}{\rulesetvar}{\type}
} \\ \\

\eda

\bda{llc}

\myruleform{\lookup{\env}{\type} = \rulet} & & 
\myirule
{
  \lookup{\rulesetvar}{\type} = \rulet \quad\quad \shade{\textsf{no\_overlap}(\rulesetvar,\type)}
}
{
  \lookup{(\env;\rulesetvar)}{\type} = \rulet
} \\ \\
& &
\myirule
{
  \lookup{\rulesetvar}{\type} = \bot \quad 
  \lookup{\env}{\type} = \rulet
}
{
  \lookup{(\env;\rulesetvar)}{\type} = \rulet
} \\ \\ 

\myruleform{\lookup{\rulesetvar}{\type} = \rulet} & &
 \myirule
{
  \rulet \in \rulesetvar \quad\quad \rulet = \forall\vec{\alpha}'.\rulesetvar'\To\type' \quad\quad
  \theta\type' = \type 
}
{
  \lookup{\rulesetvar}{\type} = \theta\rulesetvar' \Rightarrow \type
} 

\eda
}

Figure \ref{fig:type} presents the static type system
of $\ourlang$. The typing judgment 
${\tenv\mid\env\turns\relation{e}{\type}}$
means that expression $e$ has type $\type$ under type environment $\tenv$
and implicit environment $\env$. 
The auxiliary resolution judgment
$\env \vturns \rulet$
expresses that type $\rulet$ is resolvable with respect to $\env$.
Here, $\tenv$ is the conventional type environment
that captures type variables; $\env$ is the \textit{implicit environment}, defined as a stack of
contexts.  
Figure \ref{fig:type} also presents lookup in the implicit environment
($\lookup{\env}{\type}$) and in contexts
($\lookup{\rulesetvar}{\type}$). 

We will not discuss the first four rules ($\TyInt$, $\TyVar$, $\TyAbs$ and $\TyApp$) because they
are entirely standard. For now we also ignore the gray-shaded conditions in the other rules;
they are explained in Section~\ref{sec:conditions}.

Rule $\TyRule$ checks a rule abstraction $\ruleabs{\rulesch{\alpha}{\rulesetvar}{\type}}{e}$ by checking
whether the rule's body $e$ actually has the type $\type$ under the assumed
implicit type context $\bar{\rulet}$. Rule $\TyInst$ instantiates a rule
type's type variables $\vec{\alpha}$ with the given types $\vec{\type}$,
and rule $\TyRApp$ instantiates the type context $\bar{\rulet}$ with expressions
of the required rule types $\overline{\relation{e}{\rulet}}$.
Finally, rule $\TyQuery$ delegates queries directly to the resolution rule $\StaRes$.

\paragraph{Resolution Principle}
The underlying principle of resolution in $\ourlang$ originates
from resolution in logic. Following the Curry-Howard correspondence,
we assign to each type a corresponding logical interpretation
with the $(\cdot)^\dagger$ function:
\begin{definition}[Logical Interpretation]
\begin{eqnarray*}
\alpha^\dagger & = & \alpha^\dagger \\
\tyint^\dagger & = & \tyint^\dagger \\
(\type_1 \rightarrow \type_2)^\dagger & = & \type_1^\dagger \rightarrow^\dagger \type_2^\dagger \\
(\forall \vec{\alpha}.\rulesetvar \To \type)^\dagger & = & \forall \vec{\alpha}^\dagger . \bigwedge_{\rho \in \rulesetvar} \rho^\dagger \To \type^\dagger
\end{eqnarray*}
\end{definition}
Here, type variables $\alpha$ map to propositional variables $\alpha^\dagger$
and the primitive type $\tyint$ maps to the propositional constant
$\tyint^\dagger$. Unlike Curry-Howard, we do not map function types to logical
implications; we deliberately restrict our implicational reasoning to rule
types. So, instead we also map the function arrow to an uninterpreted
higher-order predicate $\rightarrow^\dagger$. Finally, as already indicated, we
map rule types to logical implications. 

Resolution in $\ourlang$ then corresponds to checking entailment of the logical
interpretation.
We postulate this property as a theorem that constrains the design of resolution.
\begin{theorem}[Resolution Specification]\label{th:resolution} \ 
\begin{center}
If $\env \vturns \rho$, then $\env^\dagger \models \rho^\dagger$.
\end{center}
\end{theorem}

\paragraph{Resolution for Simple Types}
The step from the logical interpretation to the $\StaRes$ rule in
Figure~\ref{fig:type} is non-trivial. So, let us first look
at a simpler incarnation. What does resolution look like for simple
types $\type$ like $\tyint$? 
\begin{equation*}
\SimpleRes~~~~~
\myirule
{ \lookup{\env}{\type} = \rulesetvar' \Rightarrow \type  \\
  \env \vturns \rulet_i \quad 
  (\forall \rulet_i \in \rulesetvar')
}
{ \env \vturns \type
}
~~~~~\phantom{\SimpleRes}
\end{equation*}
First, it looks up a \textit{matching} rule type in the implicit environment by
means of the lookup function $\lookup{\env}{\type}$ defined in
Fig.~\ref{fig:type}.  This partial function respects the nested scopes: it
first looks in the topmost context of the implicit environment, and, only if it
does not find a matching rule, does it descend. Within an environment context,
the lookup function looks for a rule type whose right-hand side $\type'$ can be
instantiated to the queried $\type$ using a matching unifier $\theta$. This rule
type is then returned in instantiated form.

The matching expresses that the looked-up rule produces a value of the required
type.  To do so, the looked-up rule may itself require other implicit values.
This requirement is captured in the context $\rulesetvar'$, which must
be resolved recursively. Hence, the resolution rule is
itself a recursive rule. When the context $\rulesetvar'$ of the looked-up
rule is empty, a base case of the recursion has been reached.

\begin{example}
Consider this query for a tuple of integers:  \[\tyint; \ruleschr{\alpha}{\{
\alpha \}}{\alpha \times \alpha} \vturns \tyint \times \tyint\] Lookup yields
the second rule, which produces a tuple, instantiated to $\{ \tyint \}
\Rightarrow \tyint \times \tyint$ with matching substitution $\theta = [\alpha
\mapsto \tyint]$. In order to produce a tuple, the rule requires a
value of the component type. Hence, resolution proceeds by recursively querying
for $\tyint$. Now lookup yields the first rule, which
produces an integer, with empty matching substitution and no further
requirements.
\end{example}

\paragraph{Resolution for Rule Types}
So far, so good. Apart from allowing any types, recursive querying for simple
types is quite similar to recursive type class resolution, and $\ourlang$
carefully captures the expected behavior. However, what is distinctly novel in
$\ourlang$, is that it also provides \textit{resolution of rule types}, which
requires a markedly different treatment.
\begin{equation*}
\RuleRes~~~~~
\myirule
{ \lookup{\env}{\type} = \rulesetvar \Rightarrow \type
}
{ \env \vturns \rulesch{\alpha}{\rulesetvar}{\type}
}
~~~~~\phantom{\RuleRes}
\end{equation*}
Here we retrieve a whole rule from the environment, including its context.
Resolution again performs a lookup based on a matching right-hand side $\type$,
but subsequently also matches the context with the one that is queried. No recursive
resolution takes place.
\begin{example}
Consider a variant of the above query:  
\[\tyint; \ruleschr{\alpha}{\{ \alpha \}}{\alpha \times \alpha} \vturns \{ \tyint \} \Rightarrow \tyint \times \tyint\] 
Again lookup yields the second rule, instantiated to $\{ \tyint \}
\Rightarrow \tyint \times \tyint$. The context $\{ \tyint \}$ of this rule
matches the context of the queried rule. Hence, the query is resolved without
recursive resolution.
\end{example}

\paragraph{Unified Resolution}
The feat that our actual resolution rule $\StaRes$ accomplishes is to unify
these seemingly disparate forms of resolution into one single inference rule.
In fact, both $\SimpleRes$ and $\RuleRes$ are special cases of $\StaRes$, which
provides some additional expressiveness in the form of \textit{partial
resolution} (explained below).

The first hurdle for $\StaRes$ is that types $\type$ and rule types $\rulet$
are different syntactic categories. Judging from its definition, $\StaRes$ only covers rule types.
How do we get it to treat simple types then?  Just promote the simple
type $\type$ to its corresponding rule type
$\ruleschr{}{\{\}}{\type}$ and $\StaRes$ will do what we expect
for simple types, including recursive resolution. At the same time, it still
matches proper rule types exactly, without recursion, when that is appropriate.

Choosing the right treatment for the context is the second hurdle. This part is
managed by recursively resolving $\rulesetvar' - \rulesetvar$. In the case of
promoted simple types, $\rulesetvar$ is empty, and the whole of $\rulesetvar'$
is recursively solved; which is exactly what we want. In the case
$\rulesetvar'$ matches $\rulesetvar$, no recursive resolution takes place.
Again this perfectly corresponds to what we have set out above for proper rule
types. However, there is a third case, where $\rulesetvar' - \rulesetvar$ is a
non-empty proper subset of $\rulesetvar'$. We call this situation, where part of the
retrieved rule's context is recursively resolved and part is not,
\textit{partial resolution}.

\begin{example}
Here is another query variant:  
\[\tybool; \ruleschr{\alpha}{\{ \tybool, \alpha \}}{\alpha \times \alpha}
\vturns \{ \tyint \} \Rightarrow \tyint \times \tyint\] 
The first lookup yields the second rule, instantiated to $\{ \tybool, \tyint \}
\Rightarrow \tyint \times \tyint$, which almost matches the queried rule type.
Only $\tybool$ in the context is unwelcome, so it is eliminated through a recursive
resolution step. Fortunately, the first rule in the environment is available for that.
\end{example}

\paragraph{Semantic Resolution}
Within the confines of the semantic constraint of Theorem~\ref{th:resolution}
the rule $\StaRes$ implements a rather syntactic notion of resolution.
In contrast, a fully semantic definition of resolution would coincide exactly
with the semantic constraint and satisfy
\begin{center}
$\env \vturns \rho$ iff $\env^\dagger \models \rho^\dagger$
\end{center}
For instance, it would allow to resolve
\[ \tychar ; \tychar \Rightarrow \tyint  ; \tybool \Rightarrow \tyint \vturns \tyint\] 
In this example, resolution gets stuck using the topmost rule in the
environment. However, by using the next one down, the query can be resolved.
The problem with supporting this semantic notion of resolution is that it requires
\textit{backtracking}. Because backtracking easily becomes a performance problem
and because it is mentally hard to reason about for the programmer, we have decided
against it.

We have considered another definition of resolution, that avoids backtracking
but is closer to the semantic notion:
\[
\myirule
{ \lookup{\env}{\type} = \rulesetvar' \Rightarrow \type \\
  \env,\rulesetvar \vturns \rulet_i \quad 
  (\forall \rulet_i \in \rulesetvar')
}
{ \env \vturns \rulesch{\alpha}{\rulesetvar}{\type}
} \\ \\
\]
This rule extends the environment $\env$ with the queried rule type's
context $\rulesetvar$ for recursive resolution of the matching rule type's
context $\rulesetvar'$. It resolves the following query that rule $\StaRes$ does
not:
\[ \tychar ; \tychar \Rightarrow \tyint  ; \tybool \Rightarrow \tyint \vturns \tychar \Rightarrow \tyint\] 

However, we prefer our more syntactic definition of resolution, rule $\StaRes$,
because it is much simpler: the environment does not grow recursively, but
stays the same throughout the whole recursive resolution. We believe that this
way it is more manageable for the programmer to perform resolution mentally. Moreover,
the invariant environment in rule $\StaRes$ is much easier for deciding termination.

\subsection{Additional Type System Conditions}\label{sec:conditions}

The gray-shaded conditions in the type system are to check lookup
errors ($\textsf{no\_overlap}$) 
and ambiguous
instantiations (\unambiguous). 

\paragraph{Avoiding Lookup Errors} To prevent lookup failures, we
have to check for two situations:

\begin{itemize}

\item A lookup has no matching rule in the environment.

\item A lookup has multiple matching rules which have different rule
  types but can yield values of the same type (overlapping
  rules).


\end{itemize}
The former condition is directly captured in the definition of lookup
among a set of rule types. The latter condition is captured in the
$\textsf{no\_overlap}$ property, which is defined as:
\begin{equation*}
\begin{array}{lcclcl}
 \multicolumn{6}{l}{\textsf{no\_overlap}(\{\rulet_1,\ldots,\rulet_n\},\type) 
  \defeq} \\
~~~ & \forall i, j. &        & \rulet_i = \rulesch{\alpha_i}{\rulesetvar_i}{\type_i} & \wedge & \exists \theta_i. \theta_i\type_i = \type  \\
    &               & \wedge & \rulet_j = \rulesch{\alpha_j}{\rulesetvar_j}{\type_j} & \wedge & \exists \theta_j. \theta_j\type_j = \type  \\
    & & \Longrightarrow & ~ i = j
\end{array}
\end{equation*}

\paragraph{Avoiding Ambiguous Instantiations}
We avoid ambiguous instantiations in the same way as Haskell does:
all quantified type
variables ($\vec{\alpha}$) in a rule type
($\rulesch{\alpha}{\rulesetvar}{\type}$) must occur in 
$\type$. We use the \unambiguous{} condition to check in
$\TyRule$ and $\TyQuery$:
\begin{align*}
  \unambiguous(\rulesch{\alpha}{\rulesetvar}{\type})
  =      & ~ \vec{\alpha} \subseteq \ftv{\tau} \\
  \wedge & ~ \forall
  \rulet_i \in \rulesetvar. \unambiguous(\rulet_i).
\end{align*}
If there is a quantified type variable not in type $\type$, the type
may yield ambiguous instantiations (e.g. $\forall
\alpha.\{\alpha\} \To \tyint$).

%% file: src/Translation.tex
\section{Type-Directed Translation to System F}
\label{sec:trans}

\figtwocol{f:trans}{Type-directed Translation to System F}{
\small
\bda{llrl} 
\text{Type Environments} & \tenv & ::= & \cdot \mid \tenv; \relation{x}{\type} \\
\text{Translation Environments} & \env & ::= & \cdot \mid \env; \overline{\relation{\rulet}{x}}\\
\eda \\
\bda{lc} 

\multicolumn{2}{l}{
  \myruleform{\tenv \mid \denv \turns \relation{e}{\type} \leadsto E}} \\

\TrInt &
{ \tenv \mid \denv \turns \relation{n}{\tyint} \leadsto n } 
\\ \\

\TrVar &
\myirule
{ (\relation{x}{\type}) \in \tenv}
{ \tenv \mid \denv \turns \relation{x}{\type} \leadsto x
} 
\\ \\

\TrAbs &
\myirule
{ \tenv;\relation{x}{\type_1}\mid \denv \turns \relation{e}{\type_2} \leadsto E
}
{ \tenv\mid\denv \turns \relation{\lambda \relation{x}{\type_1}.e}{\type_1 \arrow \type_2}
  \leadsto \lambda \relation{x}{|\type_1|}.E } 
\\ \\

\TrApp &
\myirule
{ \tenv\mid \denv \turns \relation{e_1}{\type_2 \arrow \type_1} \leadsto E_1 \\
  \tenv\mid \denv \turns \relation{e_2}{\type_2} \leadsto E_2
}
{ \tenv\mid\denv \turns \relation{e_1\,e_2}{\type_1} \leadsto E_1\,E_2} 
\\ \\

\TrQuery &
\myirule
{ \denv \vturns \rulet \leadsto E}
{ \tenv\mid\denv \turns \relation{?\rulet}{\rulet} \leadsto E
} 
\\ \\

\TrRule &
\myirule
{ \rulet = \rulesch{\alpha}{\ruleset}{\type} \quad
  \vec{\alpha} \cap \mathit{ftv}(\tenv,\env) = \emptyset \\
  \tenv\mid\denv; \overline{\relation{\rulet}{x}} \turns 
  \relation{e}{\type} \leadsto E \quad\quad
  \bar{x}\text{~fresh} 
}
{ \tenv\mid\denv \turns 
  \relation{\ruleabs{\rulet}{e}}{\rulet} \leadsto
  \Abs{\vec{\alpha}}{\abs{({\relation{\vec{x}}{|\vec{\rulet}|}})}{E}}
} 
\\ \\

\TrInst &
\myirule
{ \tenv\mid\denv \turns \relation{e}{\rulesch{\alpha}{\rulesetvar}{\type}} \leadsto E
}
{ \tenv\mid\denv \turns 
  \relation
  {e[\vec{\type}]}
  {\subst{\vec{\alpha}}{\vec{\type}}(\rulesetvar \To \type)} 
  \leadsto E\;|\vec{\type}|
  
} 
\\ \\

\TrRApp &
\myirule
{ \tenv\mid\denv \turns \relation{e}{\ruleset \To \type} \leadsto E \\
  \tenv\mid\denv \turns \relation{e_i}{\rulet_i} \leadsto E_i \quad 
  (\forall \relation{e_i}{\rulet_i} \in \rulesetexp)
}
{ \tenv\mid\denv \turns \relation{(\ruleapp{e}{\rulesetexp})}{\type}
  \leadsto E~\vec{E}}
\\ \\

\multicolumn{2}{l}{\myruleform{\denv \vturns \rho \leadsto E}} \\ \\

\TrRes &
\myirule
{ 
  \denv(\type) = 
  \relation{\rulesetvar' \Rightarrow \type}{E} \quad\quad
  \bar{x}\text{~fresh} \\
  
  \forall \rulet_i \in \rulesetvar': 
  \left\{
    \begin{array}{ll}
      \denv \vturns \rulet_i \leadsto E_i &
      , \rulet_i \not \in \rulesetvar  \\
      E_i = x_i &
      , \rulet_i \in \rulesetvar
    \end{array}
  \right.
}
{ \denv \vturns \rulesch{\alpha}{\rulesetvar}{\type} \leadsto
  \Abs
  {\vec{\alpha}}
  {\abs
    {(\relation{\vec{x}}{|\vec{\rulet}|})}
    {(E\,\vec{E})}}
}


\eda
\bda{llc}
\myruleform{\lookup{\env}{\type} = \rulet : E} &  
\myirule
{
  \lookup{\overline{\relation{\rulet}{x}}}{\type} = \rulet : E
}
{
  \lookup{(\env;\overline{\relation{\rulet}{x}})}{\type} = \rulet : E
} \\ \\
 & 
\myirule
{
  \lookup{\overline{\relation{\rulet}{x}}}{\type} = \bot \quad 
  \lookup{\env}{\type} = \rulet
}
{
  \lookup{(\env;\overline{\relation{\rulet}{x}})}{\type} = \rulet
} \\ \\ 

\myruleform{\lookup{\overline{\relation{\rulet}{x}}}{\type} = \rulet : E} &
 \myirule
{
  (\relation{\rulet}{x}) \in \overline{\relation{\rulet}{x}} \quad\quad \rulet = \ruleschr{\vec{\alpha}'}{\rulesetvar'}{\type'} \\
  \theta\type' = \type  \quad\quad \theta = [\vec{\alpha}' \mapsto \vec{\type} ]
}
{
  \lookup{\overline{\relation{\rulet}{x}}}{\type} = \theta\bar{\rulet}' \Rightarrow \type : x\,|\vec{\type}|
} 

\eda \\
\begin{eqnarray*}
|\alpha| & = & \alpha \\
|\tyint| & = & \tyint \\
|\type_1 \arrow \type_2| & = & |\type_1| \arrow |\type_2| \\
|\forall \vec{\alpha}.\myset{\rulet_1, \cdots, \rulet_n} \To \type| & = & 
\forall \vec{\alpha}.|\rulet_1| \to \cdots \to |\rulet_n| \to |\type| \\
|\tenv| & = & \myset{ (\relation{x}{|\type|})  ~|~ (\relation{x}{\type}) \in \tenv } \\
|\denv| & = & 
\myset{ (\relation{x}{|\rulet|}) ~|~ 
  (\relation{\rulet}{x}) \in \denv }
\end{eqnarray*}
}

In this section we define the dynamic semantics of $\ourlang$ in terms
of System F's dynamic semantics, by means of a type directed translation. 
This translation turns implicit contexts into explicit parameters and
statically resolves all queries, much like Wadler and Blott's dictionary
passing translation for type classes~\cite{adhoc}. 
The advantage of this approach is that we simultaneously provide a meaning to
well-typed $\ourlang$ programs and an effective implementation that resolves
all queries statically.

\subsection{Type-Directed Translation}
Figure~\ref{f:trans} presents the translation rules that convert $\ourlang$
expressions into ones of System F extended with the integer and unit types. 
This figure essentially extends Figure~\ref{fig:type} with the necessary
information for the translation, but for readability we have omitted the
earlier gray-shaded conditions.

The syntax of System
F is as follows: 
{\small
  \[ \begin{array}{llrl}
    \text{Types} & T & ::= & \alpha \mid T \arrow T 
    \mid \forall \alpha. T \mid \tyint \mid \tyunit \\ 
    \text{Expressions} & E & ::=  & x \mid \lambda (x:T) . E \mid E\;E
    \mid \Lambda \alpha . E \mid E\;T \mid n \mid \unit 
  \end{array} \]}

The main translation judgment is 
\begin{center}
  $\tenv \mid \denv \turns \relation{e}{\type} \leadsto E$,
\end{center}
which states that the translation of $\ourlang$ expression $e$ with
type $\type$ is System~F expression $E$, with respect to type environment
$\tenv$ and translation
environment $\denv$. The translation environment $\denv$ relates each
rule type in the earlier implicit environment to a System~F variable
$x$; this variable serves as value-level explicit evidence for the implicit rule. Lookup
in the translation environment is defined similarly to lookup in the
type environment, except that the lookup now returns a pair of a rule type
and an evidence variable.

Figure~\ref{f:trans} also defines the type translation function $|\cdot|$ from
$\ourlang$ types $\type$ to System F types T.  In order to obtain a unique
translation of types, we assume that the types in a context are
lexicographically ordered.



Variables, lambda abstractions and applications
are translated straightforwardly.
Queries are translated by rule $\TrQuery$ using the auxiliary
resolution judgment $\vturns$, defined by rule $\TrRes$. Note that
rule $\TrRes$ performs the same process that rule $\StaRes$ performs
in the type system except that it additionally collects evidence variables.

Rule $\TrRule$ translates rule abstractions to explicit type and value
abstractions in System~F, and rule $\TrInst$ translates instantiation
to type application. Finally, rule $\TrRApp$ translates rule
application to application in System~F.

\begin{example}
We have that:
\begin{gather*}
\cdot \mid \cdot \turns \ruleabs{\ruleschr{\alpha}{\{\alpha\}}{\alpha \times \alpha}}{(?{\alpha},?{\alpha})} \\
  \leadsto \Abs{\alpha}{\abs{(\relation{x}{\alpha})}{(x,x)}}
\end{gather*}
and also:
\begin{gather*}
(\relation{\tyint}{x_1}), 
(\relation{\ruleschr{\alpha}{\{\alpha\}}{\alpha \times \alpha}}{x_2})
\vturns \tyint\times\tyint \\
  \leadsto x_2\,\tyint\,x_1
\end{gather*}
\end{example}
For brevity, Figure~\ref{f:trans} omits the case where the
context of a rule type is empty. To properly handle empty contexts, the
translation of rule type should include $|\{\} \To \type| = \tyunit \to
|\type|$ and the translation rules \TrRule, \TrRApp{} and \TrRes{} should be
extended in the obvious way. 

\thmtranstypreserve
\begin{proof} (Sketch)
  We first prove\footnote{in the extra material of the submission} the more general lemma ``if $\tenv \mid \denv \turns
  \relation{e}{\type} \leadsto E$, then $|\tenv|,|\denv| \turns
  \relation{E}{|\type|}$'' by induction on the derivation of
  translation. Then, the theorem is trivially proved by it.
\end{proof}

\subsection{Dynamic Semantics}
Finally, we define the dynamic semantics of $\ourlang$ as the composition
of the type-directed translation and System F's dynamic semantics. 
Following Siek's notation~\cite{fg}, this dynamic semantics is:
\[ \mathit{eval}(e) = V \quad\quad \textit{where } \cdot\mid\cdot \turns \relation{e}{\type} \leadsto E \textit{ and } E \rightarrow^* V  \]
with $\rightarrow^*$ the reflexive, transitive closure of System F's standard single-step call-by-value reduction relation.

Now we can state the conventional type safety theorem for $\ourlang$:
\begin{theorem}[Type Safety]
If $\cdot\mid\cdot \turns \relation{e}{\type}$, then $\mathit{eval}(e) = V$ for
some System F value $V$.
\end{theorem}
The proof follows trivially from Theorem~\ref{thm:type:preservation}.

%% file: src/SourceLang.tex
%
%
\makeatletter
\@ifundefined{lhs2tex.lhs2tex.sty.read}%
  {\@namedef{lhs2tex.lhs2tex.sty.read}{}%
   \newcommand\SkipToFmtEnd{}%
   \newcommand\EndFmtInput{}%
   \long\def\SkipToFmtEnd#1\EndFmtInput{}%
  }\SkipToFmtEnd

\newcommand\ReadOnlyOnce[1]{\@ifundefined{#1}{\@namedef{#1}{}}\SkipToFmtEnd}
\usepackage{amstext}
\usepackage{amssymb}
\usepackage{stmaryrd}
\DeclareFontFamily{OT1}{cmtex}{}
\DeclareFontShape{OT1}{cmtex}{m}{n}
  {<5><6><7><8>cmtex8
   <9>cmtex9
   <10><10.95><12><14.4><17.28><20.74><24.88>cmtex10}{}
\DeclareFontShape{OT1}{cmtex}{m}{it}
  {<-> ssub * cmtt/m/it}{}
\newcommand{\texfamily}{\fontfamily{cmtex}\selectfont}
\DeclareFontShape{OT1}{cmtt}{bx}{n}
  {<5><6><7><8>cmtt8
   <9>cmbtt9
   <10><10.95><12><14.4><17.28><20.74><24.88>cmbtt10}{}
\DeclareFontShape{OT1}{cmtex}{bx}{n}
  {<-> ssub * cmtt/bx/n}{}
\newcommand{\tex}[1]{\text{\texfamily#1}}	

\newcommand{\Sp}{\hskip.33334em\relax}

\newcommand{\Conid}[1]{\mathit{#1}}
\newcommand{\Varid}[1]{\mathit{#1}}
\newcommand{\anonymous}{\kern0.06em \vbox{\hrule\@width.5em}}
\newcommand{\plus}{\mathbin{+\!\!\!+}}
\newcommand{\bind}{\mathbin{>\!\!\!>\mkern-6.7mu=}}
\newcommand{\rbind}{\mathbin{=\mkern-6.7mu<\!\!\!<}}
\newcommand{\sequ}{\mathbin{>\!\!\!>}}
\renewcommand{\leq}{\leqslant}
\renewcommand{\geq}{\geqslant}
\usepackage{polytable}

\@ifundefined{mathindent}%
  {\newdimen\mathindent\mathindent\leftmargini}%
  {}%

\def\resethooks{%
  \global\let\SaveRestoreHook\empty
  \global\let\ColumnHook\empty}
\newcommand*{\savecolumns}[1][default]%
  {\g@addto@macro\SaveRestoreHook{\savecolumns[#1]}}
\newcommand*{\restorecolumns}[1][default]%
  {\g@addto@macro\SaveRestoreHook{\restorecolumns[#1]}}
\newcommand*{\aligncolumn}[2]%
  {\g@addto@macro\ColumnHook{\column{#1}{#2}}}

\resethooks

\newcommand{\onelinecommentchars}{\quad-{}- }
\newcommand{\commentbeginchars}{\enskip\{-}
\newcommand{\commentendchars}{-\}\enskip}

\newcommand{\visiblecomments}{%
  \let\onelinecomment=\onelinecommentchars
  \let\commentbegin=\commentbeginchars
  \let\commentend=\commentendchars}

\newcommand{\invisiblecomments}{%
  \let\onelinecomment=\empty
  \let\commentbegin=\empty
  \let\commentend=\empty}

\visiblecomments

\newlength{\blanklineskip}
\setlength{\blanklineskip}{0.66084ex}

\newcommand{\hsindent}[1]{\quad}
\let\hspre\empty
\let\hspost\empty
\newcommand{\NB}{\textbf{NB}}
\newcommand{\Todo}[1]{$\langle$\textbf{To do:}~#1$\rangle$}

\EndFmtInput
\makeatother
%
%
%
%
%
%
%
%
%
\ReadOnlyOnce{polycode.fmt}%
\makeatletter

\newcommand{\hsnewpar}[1]%
  {{\parskip=0pt\parindent=0pt\par\vskip #1\noindent}}

\newcommand{\hscodestyle}{}


\newcommand{\sethscode}[1]%
  {\expandafter\let\expandafter\hscode\csname #1\endcsname
   \expandafter\let\expandafter\endhscode\csname end#1\endcsname}


\newenvironment{compathscode}%
  {\par\noindent
   \advance\leftskip\mathindent
   \hscodestyle
   \let\\=\@normalcr
   \let\hspre\(\let\hspost\)%
   \pboxed}%
  {\endpboxed\)%
   \par\noindent
   \ignorespacesafterend}

\newcommand{\compaths}{\sethscode{compathscode}}


\newenvironment{plainhscode}%
  {\hsnewpar\abovedisplayskip
   \advance\leftskip\mathindent
   \hscodestyle
   \let\hspre\(\let\hspost\)%
   \pboxed}%
  {\endpboxed%
   \hsnewpar\belowdisplayskip
   \ignorespacesafterend}

\newenvironment{oldplainhscode}%
  {\hsnewpar\abovedisplayskip
   \advance\leftskip\mathindent
   \hscodestyle
   \let\\=\@normalcr
   \(\pboxed}%
  {\endpboxed\)%
   \hsnewpar\belowdisplayskip
   \ignorespacesafterend}


\newcommand{\plainhs}{\sethscode{plainhscode}}
\newcommand{\oldplainhs}{\sethscode{oldplainhscode}}
\plainhs


\newenvironment{arrayhscode}%
  {\hsnewpar\abovedisplayskip
   \advance\leftskip\mathindent
   \hscodestyle
   \let\\=\@normalcr
   \(\parray}%
  {\endparray\)%
   \hsnewpar\belowdisplayskip
   \ignorespacesafterend}

\newcommand{\arrayhs}{\sethscode{arrayhscode}}


\newenvironment{mathhscode}%
  {\parray}{\endparray}

\newcommand{\mathhs}{\sethscode{mathhscode}}


\newenvironment{texthscode}%
  {\(\parray}{\endparray\)}

\newcommand{\texths}{\sethscode{texthscode}}


\def\codeframewidth{\arrayrulewidth}
\RequirePackage{calc}

\newenvironment{framedhscode}%
  {\parskip=\abovedisplayskip\par\noindent
   \hscodestyle
   \arrayrulewidth=\codeframewidth
   \tabular{@{}|p{\linewidth-2\arraycolsep-2\arrayrulewidth-2pt}|@{}}%
   \hline\framedhslinecorrect\\{-1.5ex}%
   \let\endoflinesave=\\
   \let\\=\@normalcr
   \(\pboxed}%
  {\endpboxed\)%
   \framedhslinecorrect\endoflinesave{.5ex}\hline
   \endtabular
   \parskip=\belowdisplayskip\par\noindent
   \ignorespacesafterend}

\newcommand{\framedhslinecorrect}[2]%
  {#1[#2]}

\newcommand{\framedhs}{\sethscode{framedhscode}}


\newenvironment{inlinehscode}%
  {\(\def\column##1##2{}%
   \let\>\undefined\let\<\undefined\let\\\undefined
   \newcommand\>[1][]{}\newcommand\<[1][]{}\newcommand\\[1][]{}%
   \def\fromto##1##2##3{##3}%
   \def\nextline{}}{\) }%

\newcommand{\inlinehs}{\sethscode{inlinehscode}}


\newenvironment{joincode}%
  {\let\orighscode=\hscode
   \let\origendhscode=\endhscode
   \def\endhscode{\def\hscode{\endgroup\def\@currenvir{hscode}\\}\begingroup}
   \orighscode\def\hscode{\endgroup\def\@currenvir{hscode}}}%
  {\origendhscode
   \global\let\hscode=\orighscode
   \global\let\endhscode=\origendhscode}%

\makeatother
\EndFmtInput
%
%
%
%
%
%
\ReadOnlyOnce{forall.fmt}%
\makeatletter


\let\HaskellResetHook\empty
\newcommand*{\AtHaskellReset}[1]{%
  \g@addto@macro\HaskellResetHook{#1}}
\newcommand*{\HaskellReset}{\HaskellResetHook}

\global\let\hsforallread\empty

\newcommand\hsforall{\global\let\hsdot=\hsperiodonce}
\newcommand*\hsperiodonce[2]{#2\global\let\hsdot=\hscompose}
\newcommand*\hscompose[2]{#1}

\AtHaskellReset{\global\let\hsdot=\hscompose}

\HaskellReset

\makeatother
\EndFmtInput

\section{Source Languages and Implicit Instantiation}
\label{sec:example}
Languages like Haskell and Scala provide a lot more programmer convenience than
$\ourlang$ (which is a low level core language) because of
higher-level GP constructs, interfaces and implicit instantiation.
This section illustrates how to build a simple source language
on top
of $\ourlang$
to add the expected convenience. 
We should note that unlike Haskell this language
supports local and nested scoping, and unlike both Haskell and Scala 
it supports higher-order rules. We present the type-directed 
translation from the source to $\ourlang$. 

\subsection{Type-directed Translation to $\ourlang$}
\label{subsec:trans_src}

\begin{figure}
\small\begin{hscode}\SaveRestoreHook
\column{B}{@{}>{\hspre}l<{\hspost}@{}}%
\column{3}{@{}>{\hspre}l<{\hspost}@{}}%
\column{8}{@{}>{\hspre}l<{\hspost}@{}}%
\column{9}{@{}>{\hspre}l<{\hspost}@{}}%
\column{25}{@{}>{\hspre}l<{\hspost}@{}}%
\column{30}{@{}>{\hspre}l<{\hspost}@{}}%
\column{56}{@{}>{\hspre}c<{\hspost}@{}}%
\column{56E}{@{}l@{}}%
\column{E}{@{}>{\hspre}l<{\hspost}@{}}%
\>[3]{}{\bf interface}\;\Conid{Eq}\;\alpha\mathrel{=}~\{\Varid{eq}\mathbin{:}\alpha\to \alpha\to \Conid{Bool}\mskip1.5mu\}{}\<[E]%
\\[\blanklineskip]%
\>[3]{}\mathbf{let}\;{}\<[8]%
\>[8]{}(\equiv )\mathbin{:}\forall \alpha\hsforall \hsdot{\circ }{.}~\{\Conid{Eq}\;\alpha\mskip1.5mu\}\Rightarrow \alpha\to \alpha\to \Conid{Bool}\mathrel{=}\Varid{eq}\mathbin{?}\mathbf{in}{}\<[E]%
\\
\>[3]{}\mathbf{let}\;{}\<[8]%
\>[8]{}\Varid{eqInt_1}\mathbin{:}\Conid{Eq}\;\Conid{Int}{}\<[25]%
\>[25]{}\mathrel{=}\Conid{Eq}\;~\{\Varid{eq}\mathrel{=}\Varid{primEqInt}\mskip1.5mu\}\;\mathbf{in}{}\<[E]%
\\
\>[3]{}\mathbf{let}\;{}\<[8]%
\>[8]{}\Varid{eqInt_2}\mathbin{:}\Conid{Eq}\;\Conid{Int}\mathrel{=}\Conid{Eq}\;~\{\Varid{eq}\mathrel{=}\lambda \Varid{x}\;\Varid{y}.\Varid{isEven}\;\Varid{x}\mathrel{\wedge}\Varid{isEven}\;\Varid{y}\mskip1.5mu\}\;\mathbf{in}{}\<[E]%
\\
\>[3]{}\mathbf{let}\;{}\<[8]%
\>[8]{}\Varid{eqBool}\mathbin{:}\Conid{Eq}\;\Conid{Bool}\mathrel{=}\Conid{Eq}\;~\{\Varid{eq}\mathrel{=}\Varid{primEqBool}\mskip1.5mu\}\;\mathbf{in}{}\<[E]%
\\
\>[3]{}\mathbf{let}\;{}\<[8]%
\>[8]{}\Varid{eqPair}\mathbin{:}\forall \alpha\hsforall \;\beta\hsdot{\circ }{.}~\{\Conid{Eq}\;\alpha,\Conid{Eq}\;\beta\mskip1.5mu\}\Rightarrow \Conid{Eq}\;(\alpha,\beta){}\<[56]%
\>[56]{}\mathrel{=}{}\<[56E]%
\\
\>[8]{}\hsindent{1}{}\<[9]%
\>[9]{}\Conid{Eq}\;~\{\Varid{eq}\mathrel{=}\lambda \Varid{x}\;\Varid{y}.{}\<[30]%
\>[30]{}\Varid{fst}\;\Varid{x}\equiv \Varid{fst}\;\Varid{y}\mathrel{\wedge}\Varid{snd}\;\Varid{x}\equiv \Varid{snd}\;\Varid{y}\mskip1.5mu\}\;\mathbf{in}{}\<[E]%
\\
\>[3]{}\mathbf{let}\;\Varid{p}_1\mathbin{:}(\Conid{Int},\Conid{Bool})\mathrel{=}(\mathrm{4},\Conid{True})\;\mathbf{in}{}\<[E]%
\\
\>[3]{}\mathbf{let}\;\Varid{p}_2\mathbin{:}(\Conid{Int},\Conid{Bool})\mathrel{=}(\mathrm{8},\Conid{True})\;\mathbf{in}{}\<[E]%
\\
\>[3]{}{\bf implicit}\;~\{\Varid{eqInt_1},\Varid{eqBool},\Varid{eqPair}\mskip1.5mu\}\;\mathbf{in}{}\<[E]%
\\[\blanklineskip]%
\>[3]{}(\Varid{p}_1\equiv \Varid{p}_2,{\bf implicit}\;~\{\Varid{eqInt_2}\mskip1.5mu\}\;\mathbf{in}\;\Varid{p}_1\equiv \Varid{p}_2){}\<[E]%
\ColumnHook
\end{hscode}\resethooks
\caption{Encoding the Equality Type Class}

\label{fig:eq}

\end{figure}

\newcommand{\typeSrc}{T}
\newcommand{\envSrc}{G}
\figtwocol{f:syntax}{Syntax of Source Language}{
\small
\bda{l}

\ba{l}
\textbf{Interface Declarations} \\
\quad \textbf{interface}~I~\vec{\alpha}~=~\overline{\relation{u}{T}}
\ea 
\\ \\

\ba{lrll}
\multicolumn{3}{l}{\textbf{Types}} \\
\typeSrc & ::=  & \alpha             & \mbox{Type Variables} \\
      & \mid & \tyint                & \mbox{Integer Type} \\
      & \mid & I~\vec{\typeSrc}       & \mbox{Interface Type} \\
      & \mid & \typeSrc \to \typeSrc    & \mbox{Function} \\

\sigma & ::=  & \forall \overline{\alpha}.~\overline{\sigma} \To \typeSrc & \mbox{Rule Type} \\

\ea 
\\ \\

\ba{lrll}
\multicolumn{3}{l}{\textbf{Expressions}} \\
E & ::=  & n & \mbox{Integer Literal} \\
  & \mid & x & \mbox{Lambda Variable} \\
  & \mid & \abs{x}{E} & \mbox{Abstraction} \\
  & \mid & E_1~E_2 & \mbox{Application} \\
  & \mid & u & \mbox{Let Variable} \\
  & \mid & {\bf let~} u~$: $ \sigma ~=~ E_1 {\bf~in~} E_2 & \mbox{Let}\\
  & \mid & {\bf implicit~} \overline{u} {\bf~in~} E_2 & \mbox{Implicit Scoping} \\
  & \mid & ? & \mbox{Implicit Lookup} \\
  & \mid & I~\overline{u = E} & \mbox{Interface Implementation}
\ea

\eda
}

\newcommand{\sem}[1]{\llbracket#1\rrbracket}

\figtwocol{f:type}{Type-directed Encoding of Source Language in $\ourlang$}{
\small
\bda{llrl} 
\text{Type Environments} & \envSrc & ::= & \cdot \mid \envSrc, \relation{u}{\sigma} \mid \envSrc, \relation{x}{\typeSrc} 
\eda 
\bda{lc} 
\multicolumn{2}{l}{\myruleform{\envSrc \turns \relation{E}{\typeSrc} \leadsto e}} \\ \\

\TyIntL &
{ \envSrc \turns n : \tyint \leadsto n } \\ \\

\TyVar &
\myirule
{ \envSrc(x) = \typeSrc }
{ \envSrc \turns \relation{x}{\typeSrc} \leadsto x } 
\\ \\


\TyAbs &
\myirule
{ \envSrc, \relation{x}{\typeSrc_1} \turns 
  E \leadsto e 
}
{ \envSrc \turns 
  \relation{\abs{x}{E}}{\typeSrc_1 \to \typeSrc_2} \leadsto \abs{\relation{x}{\sem{\typeSrc_1}}}{e}
} 
\\ \\

\TyApp &
\myirule
{ \envSrc \turns \relation{E_1}{\typeSrc_1 \to \typeSrc_2} \leadsto e_1 \\
  \envSrc \turns \relation{E_2}{\typeSrc_1} \leadsto e_2
}
{ \envSrc \turns \relation{E_1~E_2}{\typeSrc_2} \leadsto e_1 \, e_2}
\\ \\

\TyLVar &
\myirule
{ \envSrc(u) = \forall \overline{\alpha}.~\overline{\sigma} \To \typeSrc' \\
 \theta = [\overline{\alpha} \mapsto \overline{\typeSrc}]\quad \typeSrc = \theta \typeSrc' \\
 q_i = (?{\sem{\theta \sigma_i}}):{\sem{\theta \sigma_i}} \quad (\forall \sigma_i \in \overline{\sigma})
}
{ \envSrc \turns u : \typeSrc \leadsto u[\sem{\vec{\typeSrc}}] {\bf~with~} \overline{q}} 
\\ \\

\TyLet &
\myirule
{ \sigma = \forall \overline{\alpha}. \overline{\sigma} \To \typeSrc_1 \\
  \envSrc \turns \relation{E_1}{\typeSrc_1} \leadsto e_1 \\
  \envSrc, \relation{u}{\sigma} \turns \relation{E_2}{\typeSrc_2} \leadsto e_2
}
{ \envSrc \turns 
  \relation{{\bf~let~} \relation{u}{\sigma} = E_1 {\bf~in~} E_2}{\typeSrc_2}  \leadsto \\ 
  (\abs{\relation{u}{\sem{\sigma}}}{e_2})~\ruleabs{\sem{\sigma}}{e_1}
} 
\\ \\

\TyImp &
\myirule
{ \envSrc \turns \relation{E}{\typeSrc} \leadsto e \\
  \envSrc(u_i) = \sigma_i \quad 
  q_i = \relation{u_i}{\sem{\sigma_i}} \quad
  (\forall u_i \in \overline{u})
}
{ \envSrc \turns 
  \relation{{\bf~implicit~} \overline{u} {\bf~in~} E}{\typeSrc} \leadsto \\ 
  \ruleabs{\llbracket\overline{\sigma}\rrbracket \To \sem{\typeSrc}}{e} {\bf~with~} \overline{q}
} 
\\ \\

\TyIVar &
{ \envSrc \turns
  \relation{?}{\typeSrc} \leadsto  ~?(\{\} \To \sem{\typeSrc}) {\bf~with~} \{\}
} 
\\ \\

\TyRec &
\myirule
{ \forall i: \left\{\begin{array}{l}
             G(u_i) = \forall \bar{\alpha}.\{\} \Rightarrow I~\vec{\alpha} \rightarrow T_i \\
             \envSrc \turns E_i : \theta T_i \leadsto e \quad \theta = [\vec{\alpha} \mapsto \vec{T}]
             \end{array}\right.}
{ \envSrc \turns
  \relation{I~\overline{u = E}}{I~\vec{T}} \leadsto I~\overline{u = e}
} 
\eda
\bda{rcl}
\sem{\alpha} & = & \alpha \\
\sem{\tyint} & = & \tyint \\
\sem{\typeSrc_1 \to \typeSrc_2} & = & \sem{\typeSrc_1} \to \sem{\typeSrc_2} \\
\sem{I~\vec{\typeSrc}} & = & I~\sem{\vec{\typeSrc}} \\
\sem{\forall \overline{\alpha}. \overline{\sigma} \To \typeSrc} & = & \forall \sem{\vec{\alpha}}. \sem{\overline{\sigma}} \To \sem{\typeSrc}\\
\eda

}

The full syntax is presented in Figure~\ref{f:syntax}. Its use is illustrated
in the program of Figure~\ref{fig:eq}, which comprises an encoding of Haskell's
equality type class \texttt{Eq}.
The example shows that the source language features a simple type of interface $I~\vec{T}$
(basically records), which are used to encode simple forms of type classes. 
Note that we follow Haskell's conventions for records: field 
names \ensuremath{\Varid{u}} are unique and they are modeled as regular functions taking a 
record as the first argument. So a field $u$ with type $T$ in an interface
declaration $I~\vec{\alpha}$ actually has type $\forall \bar{\alpha}.\{\} \To I~\vec{\alpha} \rightarrow T$.
There are also other conventional programming constructs (such as \ensuremath{\mathbf{let}}
expressions, lambdas and primitive types). 

Unlike the core language, we strongly differentiate between simple types $T$
and type schemes $\sigma$ in order to facilitate type inference. Moreover, as
the source language provides implicit rather than explicit type instantiation,
the order of type variables in a quantifier is no longer relevant. Hence, they
are represented by a set ($\forall \bar{\alpha}$). We also distinguish simply
typed variables $x$ from let-bound variables $u$ with polymorphic type
$\sigma$.

Figure~\ref{f:type} presents the type-directed translation $G \turns E : T
\leadsto e$ of source language expressions $E$ of type $T$ to core expressions
$e$, with respect to type environment $G$. The type environment collects both
simply and polymorphic variable typings. The connection between source types
$T$ and $\sigma$ on the one hand and core types $\type$ and $\rulet$ on the
other hand is captured in the auxiliary function $\sem{\cdot}$. Note that this
function imposes a canonical ordering $\vec{\alpha}$ on the set of quantifier variables
$\bar{\alpha}$ (based on their precedence in the left-to-right prefix traversal
of the quantified type term).
For the translation of records, we assume that $\ourlang$ is extended
likewise with records.

\paragraph{\ensuremath{\mathbf{let}} and \ensuremath{\mathbf{let}}-bound variables}

The
rule $\TyLet$ in Figure~\ref{f:type} shows the type-directed translation for \ensuremath{\mathbf{let}}
expressions. This translation binds the variable \ensuremath{\Varid{u}} using a regular
lambda abstraction in an expression \ensuremath{e_{2}}, which is the result of the 
translation of the body of the \ensuremath{\mathbf{let}} construct (\ensuremath{E_{2}}). Then it applies
that abstraction to a rule whose rule type is just the corresponding  
(translated) type of the definition ($\sigma_1$), and whose body is 
the translation of the expression \ensuremath{E_{1}}.

The source language  provides convenience to the user by inferring type
arguments and implicit values automatically. This inference happens in rule
$\TyLVar$, i.e., the use of \ensuremath{\mathbf{let}}-bound variables.  That rule recovers the type
scheme of variable \ensuremath{\Varid{u}} from the environment \ensuremath{\Conid{G}}. Then it
instantiates the type scheme and fires the necessary queries to resolve the context. 

\paragraph{Queries} The source language also 
includes a query operator  (\ensuremath{\mathbin{?}}). Unlike $\ourlang$ this
query operator does not explicitly state the type; that information is provided
implicitly through type inference. 
For example, instead of using \ensuremath{\Varid{p}_1\equiv \Varid{p}_2} in Figure~\ref{f:type}, we
could have directly used the field \ensuremath{\Varid{eq}} as follows:
\begin{hscode}\SaveRestoreHook
\column{B}{@{}>{\hspre}l<{\hspost}@{}}%
\column{3}{@{}>{\hspre}l<{\hspost}@{}}%
\column{E}{@{}>{\hspre}l<{\hspost}@{}}%
\>[3]{}\Varid{eq}\mathbin{?}\Varid{p}_1\;\Varid{p}_2{}\<[E]%
\ColumnHook
\end{hscode}\resethooks
When used in this way, the query acts like a Coq placeholder (\ensuremath{\anonymous }),
which similarly instructs Coq to automatically infer a value. 


The translation  of source language queries, given by the rule
$\TyIVar$, is fairly straightforward. To simplify type-inference, the
query is limited to types, and does not support partial resolution 
(although other designs with more powerful queries are possible).
In the translated code the query is combined with a rule instantiation
and application in order to eliminate the empty rule set.

\paragraph{Implicit scoping} The \ensuremath{{\bf implicit}} construct, which has been 
already informally introduced in Section~\ref{sec:intro}, is the core scoping 
construct of the source language. It is used in our example to
first introduce definitions in the implicit environment (\ensuremath{\Varid{eqInt_1}}, 
\ensuremath{\Varid{eqBool}} and \ensuremath{\Varid{eqPair}}) available at the expression
\begin{hscode}\SaveRestoreHook
\column{B}{@{}>{\hspre}l<{\hspost}@{}}%
\column{3}{@{}>{\hspre}l<{\hspost}@{}}%
\column{E}{@{}>{\hspre}l<{\hspost}@{}}%
\>[3]{}(\Varid{p}_1\equiv \Varid{p}_2,{\bf implicit}\;~\{\Varid{eqInt_2}\mskip1.5mu\}\;\mathbf{in}\;\Varid{p}_1\equiv \Varid{p}_2){}\<[E]%
\ColumnHook
\end{hscode}\resethooks
Within this expression there is a second occurrence of \ensuremath{{\bf implicit}}, 
which introduces an overlapping rule (\ensuremath{\Varid{eqInt_2}}) that takes
priority over \ensuremath{\Varid{eqInt_1}} for the subexpression \ensuremath{\Varid{p}_1\equiv \Varid{p}_2}.

The translation rule $\TyImp$ of \ensuremath{{\bf implicit}} into $\ourlang$ also exploits
type-information to avoid redundant type annotations. For 
example, it is not necessary to annotate the \ensuremath{\mathbf{let}}-bound variables 
used in the rule set $\overline{u}$ because that information can 
be recovered from the environment \ensuremath{\Conid{G}}.

\paragraph{Higher-order rules and implicit instantiation for any type} 
The following example illustrates higher-order
rules and implicit instantiation working for any type in the source language.

\begin{hscode}\SaveRestoreHook
\column{B}{@{}>{\hspre}l<{\hspost}@{}}%
\column{3}{@{}>{\hspre}l<{\hspost}@{}}%
\column{12}{@{}>{\hspre}l<{\hspost}@{}}%
\column{14}{@{}>{\hspre}l<{\hspost}@{}}%
\column{E}{@{}>{\hspre}l<{\hspost}@{}}%
\>[3]{}\mathbf{let}\;\Varid{show}\mathbin{:}\forall \alpha\hsforall \hsdot{\circ }{.}~\{\alpha\to \Conid{String}\mskip1.5mu\}\Rightarrow \alpha\to \Conid{String}\mathrel{=}~?~\;\mathbf{in}{}\<[E]%
\\
\>[3]{}\mathbf{let}\;\Varid{showInt}\mathbin{:}\Conid{Int}\to \Conid{String}\mathrel{=}\ldots\;\mathbf{in}{}\<[E]%
\\
\>[3]{}\mathbf{let}\;\Varid{comma}\mathbin{:}\forall \alpha\hsforall \hsdot{\circ }{.}~\{\alpha\to \Conid{String}\mskip1.5mu\}\Rightarrow \![\alpha\mskip1.5mu]\to \Conid{String}\mathrel{=}\ldots\;\mathbf{in}{}\<[E]%
\\
\>[3]{}\mathbf{let}\;\Varid{space}\mathbin{:}\forall \alpha\hsforall \hsdot{\circ }{.}~\{\alpha\to \Conid{String}\mskip1.5mu\}\Rightarrow \![\alpha\mskip1.5mu]\to \Conid{String}\mathrel{=}\ldots\;\mathbf{in}{}\<[E]%
\\
\>[3]{}\mathbf{let}\;\Varid{o}\mathbin{:}{}\<[12]%
\>[12]{}~\{\Conid{Int}\to \Conid{String},~\{\Conid{Int}\to \Conid{String}\mskip1.5mu\}\Rightarrow \![\Conid{Int}\mskip1.5mu]\to \Conid{String}\mskip1.5mu\}{}\<[E]%
\\
\>[12]{}\hsindent{2}{}\<[14]%
\>[14]{}\Rightarrow \Conid{String}\mathrel{=}\Varid{show}\;\![\mathrm{1},\mathrm{2},\mathrm{3}\mskip1.5mu]\;\mathbf{in}{}\<[E]%
\\
\>[3]{}{\bf implicit}\;\Varid{showInt}\;\mathbf{in}{}\<[E]%
\\
\>[3]{}({\bf implicit}\;\Varid{comma}\;\mathbf{in}\;\Varid{o},{\bf implicit}\;\Varid{space}\;\mathbf{in}\;\Varid{o}){}\<[E]%
\ColumnHook
\end{hscode}\resethooks
For brevity, we have omitted the implementations of \ensuremath{\Varid{showInt}}, \ensuremath{\Varid{comma}} and \ensuremath{\Varid{space}};
but \ensuremath{\Varid{showInt}} renders an \ensuremath{\Conid{Int}} as a \ensuremath{\Conid{String}} in the conventional way, while
\ensuremath{\Varid{comma}} and \ensuremath{\Varid{space}} provide two ways for rendering lists.  Evaluation of
the expression yields \ensuremath{(\text{\tt \char34 1,2,3\char34},\text{\tt \char34 1~2~3\char34})}. Thanks to the implicit rule
parameters, the contexts of the two calls to \ensuremath{\Varid{o}} control how the lists
are rendered.

This example differs from that in Figure~\ref{fig:eq} in that instead of 
using a \emph{nominal} interface type like \ensuremath{\Conid{Eq}}, it uses
standard functions to model a simple concept for pretty printing values. 
The use of functions as implicit values leads to a programming style akin
to \emph{structural} matching of concepts, since only the type of the function 
matters for resolution. 

%
%
%
%
%
%
%



\subsection{Extensions}\label{subsec:extensions}

The goal of our work is to present a minimal and general framework for 
implicits. As such we have avoided making assumptions about extensions 
that would be useful for some languages, but not others. 

In this section we briefly discuss some extensions that would be useful in
the context of particular languages and the implications that they
would have in our framework.

\paragraph{Full-blown Concepts} The most noticeable feature that was not
discussed is a full-blown notion of concepts. One reason not to commit
to a particular notion of concepts is that there is no general
agreement on what the right notion of concepts is.  For example, following Haskell type classes,
the C++0x concept proposal~\cite{concepts} is based
on a \textit{nominal} approach with \textit{explicit} concept refinement, while
Stroustrup favors a \textit{structural} approach with \textit{implicit} concept refinement
because that would be more familiar to C++
programmers~\cite{stroustrup09concepts}. Moreover, various other proposals for
GP mechanisms have their own notion of interface: Scala uses standard OO
hierarchies; Dreyer et al. use ML-modules~\cite{modular}; and in
dependently typed systems (dependent) record types are used~\cite{coqclasses,instanceargs}.

An advantage of $\ourlang$ is that no particular notion of
interface is imposed on source language designers. Instead, language
designers are free to use the one they prefer. In
our source language, for simplicity, we opted to add a very simple
(and limited) type of interface. But existing language
designs~\cite{implicits,modular,coqclasses,instanceargs} offer
evidence that more sophisticated types of interfaces, including some
form of refinement or associated types, can be built on top of
$\ourlang$.

\paragraph{Type Constructor Polymorphism and Higher-order Rules}
Type constructor polymorphism is an advanced, but highly powerful GP feature
available in Haskell and Scala, among others.  It allows abstracting container
types like \ensuremath{\Conid{List}} and \ensuremath{\Conid{Tree}} with a type variable \ensuremath{\Varid{f}}; and applying the
abstracted container type to different element types, e.g., \ensuremath{\Varid{f}\;\Conid{Int}} and \ensuremath{\Varid{f}\;\Conid{Bool}}.

This type constructor polymorhism leads to a need for higher-order rules: rules
for containers of elements that depend on rules for the elements. The 
instance for showing values of type \ensuremath{\Conid{Perfect}\;\Varid{f}\;\alpha} in Section~\ref{sec:intro}, is a typical example of this need. 

Extending $\ourlang$ with type constructor polymorphism is not hard. 
Basically, we need to add a kind system and move from a System~$F$ like 
language to System $F_{\omega}$ like language.

\paragraph{Subtyping} Languages like Scala or C++ have subtyping. Subtyping
would require significant adaptations to $\ourlang$.  Essentially, instead of
targetting System F, we would have to target a version of System F with
subtyping. In addition, the notion of matching in the lookup function
$\lookup{\env}{\type}$ would have to be adjusted, as well as the
$\textit{no\_overlap}$ condition. While subtyping is a useful feature, some
language designs do not support it because it makes the system more complex and
interferes with type-inference.  

\paragraph{Type-inference}
Languages without subtyping (like Haskell or ML) make it easier to
support better type-inference. Since we do not use subtyping, 
it is possible to improve support for type-inference in our
source language. In particular, we currently require a type annotation 
for \ensuremath{\mathbf{let}} expressions, but it should be possible to make that
annotation optional, by building on existing work 
for the GHC Haskell compiler~\cite{schrijvers:outsidein,vytiniotis:outsidein}.


%% file: src/Related.tex
%
%
\makeatletter
\@ifundefined{lhs2tex.lhs2tex.sty.read}%
  {\@namedef{lhs2tex.lhs2tex.sty.read}{}%
   \newcommand\SkipToFmtEnd{}%
   \newcommand\EndFmtInput{}%
   \long\def\SkipToFmtEnd#1\EndFmtInput{}%
  }\SkipToFmtEnd

\newcommand\ReadOnlyOnce[1]{\@ifundefined{#1}{\@namedef{#1}{}}\SkipToFmtEnd}
\usepackage{amstext}
\usepackage{amssymb}
\usepackage{stmaryrd}
\DeclareFontFamily{OT1}{cmtex}{}
\DeclareFontShape{OT1}{cmtex}{m}{n}
  {<5><6><7><8>cmtex8
   <9>cmtex9
   <10><10.95><12><14.4><17.28><20.74><24.88>cmtex10}{}
\DeclareFontShape{OT1}{cmtex}{m}{it}
  {<-> ssub * cmtt/m/it}{}
\newcommand{\texfamily}{\fontfamily{cmtex}\selectfont}
\DeclareFontShape{OT1}{cmtt}{bx}{n}
  {<5><6><7><8>cmtt8
   <9>cmbtt9
   <10><10.95><12><14.4><17.28><20.74><24.88>cmbtt10}{}
\DeclareFontShape{OT1}{cmtex}{bx}{n}
  {<-> ssub * cmtt/bx/n}{}
\newcommand{\tex}[1]{\text{\texfamily#1}}	

\newcommand{\Sp}{\hskip.33334em\relax}

\newcommand{\Conid}[1]{\mathit{#1}}
\newcommand{\Varid}[1]{\mathit{#1}}
\newcommand{\anonymous}{\kern0.06em \vbox{\hrule\@width.5em}}
\newcommand{\plus}{\mathbin{+\!\!\!+}}
\newcommand{\bind}{\mathbin{>\!\!\!>\mkern-6.7mu=}}
\newcommand{\rbind}{\mathbin{=\mkern-6.7mu<\!\!\!<}}
\newcommand{\sequ}{\mathbin{>\!\!\!>}}
\renewcommand{\leq}{\leqslant}
\renewcommand{\geq}{\geqslant}
\usepackage{polytable}

\@ifundefined{mathindent}%
  {\newdimen\mathindent\mathindent\leftmargini}%
  {}%

\def\resethooks{%
  \global\let\SaveRestoreHook\empty
  \global\let\ColumnHook\empty}
\newcommand*{\savecolumns}[1][default]%
  {\g@addto@macro\SaveRestoreHook{\savecolumns[#1]}}
\newcommand*{\restorecolumns}[1][default]%
  {\g@addto@macro\SaveRestoreHook{\restorecolumns[#1]}}
\newcommand*{\aligncolumn}[2]%
  {\g@addto@macro\ColumnHook{\column{#1}{#2}}}

\resethooks

\newcommand{\onelinecommentchars}{\quad-{}- }
\newcommand{\commentbeginchars}{\enskip\{-}
\newcommand{\commentendchars}{-\}\enskip}

\newcommand{\visiblecomments}{%
  \let\onelinecomment=\onelinecommentchars
  \let\commentbegin=\commentbeginchars
  \let\commentend=\commentendchars}

\newcommand{\invisiblecomments}{%
  \let\onelinecomment=\empty
  \let\commentbegin=\empty
  \let\commentend=\empty}

\visiblecomments

\newlength{\blanklineskip}
\setlength{\blanklineskip}{0.66084ex}

\newcommand{\hsindent}[1]{\quad}
\let\hspre\empty
\let\hspost\empty
\newcommand{\NB}{\textbf{NB}}
\newcommand{\Todo}[1]{$\langle$\textbf{To do:}~#1$\rangle$}

\EndFmtInput
\makeatother
%

\section{Related Work}
\label{sec:related}

%
Throughout the paper we have already discussed a lot of related work. 
In what follows, we offer a more
detailed technical comparison of $\ourlang$ versus System $F^{G}$ and Scala 
implicits, which are the closest to our work. Then we discuss the
relation with other work in the literature.

\paragraph{System $F^{G}$} Generally speaking our calculus is more
primitive and general than System $F^{G}$. 
In contrast to $\ourlang$, System $F^{G}$ has both a notion of concepts
and implicit instantiation of concepts\footnote{Note
  that instantiation of type variables is still explicit.}. 
This has the advantage that language designers can just reuse
that infrastructure, instead of having to implement it. The language 
G~\cite{G} is based on System $F^{G}$ and it makes good use of these built-in mechanisms. However, 
System $F^{G}$ also imposes important design choices. Firstly it forces the
language designer to use the notion of concepts that is built-in to
System $F^{G}$. In contrast $\ourlang$ offers a freedom of choice (see
also the discussion in Section~\ref{subsec:extensions}). Secondly, fixing implicit
instantiation in the core prevents useful alternatives. For example, Scala and several other
systems do provide implicit instantiation by default, but also offer the option of explicit
instantiation, which is useful to resolve
ambiguities~\cite{implicits,named_instance,implicit_explicit,modular}.  This cannot be modeled
on top of System~$F^{G}$, because explicit instantiation is not
available. In contrast, by taking explicit instantiation (rule
application) as a core feature, $\ourlang$ can serve as a target for languages that offer 
both styles of instantiation.
 
There are also important differences in terms of scoping and
resolution of rules.  System $F^{G}$ only formalizes a very simple
type of resolution, which does not support recursive resolution. 
Furthermore, scoping is less fine-grained than in
$\ourlang$. For example, System $F^{G}$ requires a built-in 
construct for \ensuremath{{\bf model}} expressions, but in $\ourlang$ \ensuremath{\bf{implicit}}
(which plays a similar role) is just syntactic sugar on top of 
more primitive constructs.


\paragraph{Scala Implicits}

Scala implicits are integrated in a full-blown language, but they have
only been informally described in the literature~\cite{implicits,scala}. Our calculus aims at
providing a formal model of implicits, but there are some noteworthy
differences between $\ourlang$ and Scala implicits. In contrast to
$\ourlang$, Scala has subtyping. As discussed in Section~\ref{subsec:extensions}
subtyping would require some adaptations to our calculus. 
In Scala, nested scoping can only happen through subclassing and the rules for
resolution in the presence of overlapping instances are quite
ad-hoc. Furthermore, Scala has no (first-class) rule abstractions. Rather,
implicit arguments can only be used in definitions. In contrast
$\ourlang$ provides a more general and disciplined account of scoping
for rules.

\paragraph{Type Classes}
Obviously, the original work on type classes~\cite{adhoc} and the framework of
\textit{qualified types}~\cite{simpl_qual} around it has greatly influenced our own work,
as well as that of System F$^G$ and Scala.

There is a lot of work on Haskell type classes in the
literature.  Notably, there have been some proposals for addressing
the limitations that arise from global
scoping~\cite{named_instance,implicit_explicit}.  However in
those designs, type classes are still second-class and resolution only
works for type classes. The GHC Haskell compiler supports overlapping
instances~\cite{designspace}, that live in the same global scope. This
allows some relief for the lack of local scoping.
A lot of recent work on type classes is
focused on increasingly more powerful ``type class'' interfaces.  \emph{Functional dependencies}~\cite{fundeps},
\emph{associated types}~\cite{assoctypes,assoctypes2} and \emph{type
  families}~\cite{typefunc} are all examples of this trend.  This line
of work is orthogonal to our work.

\paragraph{Other Languages and Systems} \textit{Modular type
classes}~\cite{modular} are a language design that uses ML-modules 
to model type classes. The main novelty of this design is that, in addition
to explicit instantiation of modules, implicit instantiation is also 
supported. In contrast to $\ourlang$, implicit instantiation is
limited to modules and, although local scoping is allowed, it cannot
be nested. 

\emph{Instance arguments}~\cite{instanceargs} are an Agda extension
that is closely related to implicits.  However, unlike most GP mechanisms,
implicit rules are not declared explicitly.
Furthermore resolution is limited in its expressive power, to avoid
introducing a different computational model in Agda. This design
differs significantly from $\ourlang$, where resolution is very
expressive and the scoping mechanisms allow explicit rule declarations.

\emph{Implicit parameters}~\cite{implicit_param} are a Haskell extension that
allows \emph{named} arguments to be passed implicitly. Implicit
parameters are resolved by name, not by type and there is no 
recursive resolution.

\paragraph{GP and Logic Programming}
The connection between Haskell type classes and Prolog is folklore. 
Neubauer et. al.~\cite{Neubauer} also explore the connection with \textit{Functional Logic Programming}
and consider different evaluation strategies to deal with overlapping rules.
With \emph{Constraint Handling Rules}, Stuckey and Sulzmann~\cite{theory_over} use \textit{Constraint Logic Programming}
to implement type classes.


%% file: src/Conclusion.tex
\section{Conclusion}
\label{sec:conclusion}

Our main contribution is the development of the implicit
calculus $\ourlang$. This calculus isolates and formalizes the key
ideas of Scala implicits and provides a simple model for language designers 
interested in developing similar mechanisms for their own languages. 
In addition, $\ourlang$ supports higher-order rules and partial resolution, 
which add considerable expressiveness to the calculus.

Implicits provide an interesting alternative to conventional GP 
mechanisms like type classes or concepts. By decoupling resolution 
from a particular type of interfaces, implicits make resolution 
more powerful and general. Furthermore, this decoupling has other benefits too. 
For example, by modeling concept interfaces as conventional types, those interfaces can 
be abstracted as any other types, avoiding the issue of second class interfaces 
that arise with type classes or concepts. 

Ultimately, all the expressiveness offered by $\ourlang$
offers a wide-range of possibilities for new generic programming applications.

%% file: src/Termination.tex
\section{Termination of Resolution}

If we are not careful about which rules are made implicit, the recursive
resolution process may not terminate. This section describes how to impose 
a set of modular syntactic restrictions that prevents non-termination. 

As an example of non-termination consider 
\begin{equation*}
\myset{
  \myset{\tychar} \To \tyint,
  \myset{\tyint} \To \tychar} \vturns \tyint
\end{equation*}
which loops, using alternatively the first and second rule in the implicit
environment. 

The problem of non-termination has been widely studied in the context of
Haskell's type classes, and a set of modular syntactic restrictions
has been imposed on type class instances to avoid non-termination~\cite{fdchr}. 
Adapting these restrictions to our setting, we obtain the following termination
condition.

\defterm

%

%% file: src/Proofs.tex
\figtwocol{fig:ftype}{System F Type System}{
\bda{lc}
  (\texttt{F-Int}) & 
 \Gamma \turns n : Int \\ \\

  (\texttt{F-Var}) & 
\myirule{
           (x : T) \in \Gamma
 }{
            \Gamma \turns x : T
} \\ \\

  (\texttt{F-Abs}) & 
\myirule{
           \Gamma, x : T_1 \turns E : T_2
 }{
           \Gamma \turns \lambda x:T_1.E : T_1 \rightarrow T_2
} \\ \\

  (\texttt{F-App}) & 
\myirule{
  \Gamma \turns E_1 : T_2 \rightarrow T_1 \\
           \Gamma \turns E_2 : T_2
          }{
           \Gamma \turns E_1 \, E_2 : T_1
} \\ \\

  (\texttt{F-TApp}) & 
\myirule{
  \Gamma \turns E : \forall \alpha. T_2
           }{
            \Gamma \turns E \, T_1 : [\alpha \mapsto T_1] T_2
} \\ \\

  (\texttt{F-TAbs}) & 
\myirule{
   \Gamma \turns E : T \quad \alpha \not\in \mathit{ftv}(\Gamma)
            }{
             \Gamma \turns \Lambda \alpha.E : \forall \alpha. T 
} \\ \\
\eda
}

\section{Proofs}

Throughout the proofs we refer to the type system rules of System F listed
in Figure~\ref{fig:ftype}.

{\centering
\fbox{
\begin{minipage}{0.95\columnwidth}
\begin{lemma}
  If 
\begin{equation*}
    \tenv | \env \turns e : \type \leadsto E
\end{equation*}
  then
\begin{equation*}
    |\tenv|,|\env| \turns E : |\type|
\end{equation*}
\end{lemma}
\end{minipage}
}}

\begin{proof}
By structural induction on the expression and corresponding inference rule.
\begin{description}
\renewcommand{\itemsep}{10mm}
\item[(\texttt{TrInt})\quad$\tenv | \env \turns n : \tyint \leadsto n$] \ \\

  It follows trivially from (\texttt{F-Int}) that
\begin{equation*} 
    |\tenv|, |\env| \turns n : \tyint
\end{equation*} 

\item[(\texttt{TrVar})\quad$\tenv | \env \turns x : \type \leadsto x$] \ \\

 It follows from (\texttt{TrVar}) that 
\begin{equation*} 
    (x : \type) \in \tenv
\end{equation*} 

 Based on the definition of $|\cdot|$  it follows 
\begin{equation*} 
   (x : |\type|) \in |\tenv| 
\end{equation*} 

 Thus we have by (\texttt{F-Var}) that
\begin{equation*} 
   |\tenv|,|\env| \turns x : |\type|
\end{equation*} 

\item[(\texttt{TrAbs})\quad$\tenv | \env \turns \lambda x:\type_1.e : \type_1 \rightarrow \type_2 \leadsto \lambda x:|\type_1|.E$] \ \\

  It follows from (\texttt{TrAbs}) that
\begin{equation*} 
    \tenv ; x : \type_1 | \env \turns e : \type_2 \leadsto E
\end{equation*} 

  and by the indution hypothesis that
\begin{equation*} 
    |\tenv|, x : |\type_1|, |\env| \turns E : |\type_2|
\end{equation*} 

  As all variables are renamed unique, it is easy to verify that this also holds:
\begin{equation*} 
    |\tenv|, |\env|, x : |\type_1| \turns E : |\type_2|
\end{equation*} 

  Hence, by (\texttt{F-Abs}) we have
\begin{equation*} 
    |\tenv|, |\env| \turns \lambda x:|\type_1|.E : |\type_1 \rightarrow \type_2|
\end{equation*} 

\item[(\texttt{TrApp})\quad$\tenv | \env \turns e_1\,e_2 : \type_1 \leadsto E_1\,E_2$] \ \\

  By the induction hypothesis, we have:
\begin{equation*} 
   |\tenv|, |\env| \turns E_1 : |\type_2 \rightarrow \type_1|
\end{equation*} 

  and
\begin{equation*} 
   |\tenv|, |\env| \turns E_2 : |\type_2|
\end{equation*} 

  Then it follows by (\texttt{F-App}) that
\begin{equation*} 
   |\tenv|, |\env| \turns E_1\, E_2 : |\type_1|
\end{equation*} 

\item[(\texttt{TrQuery})\quad$\tenv | \env \turns ?\rulet : \rulet \leadsto E$] \ \\

  From (\texttt{TrQuery}) we have
\begin{equation*} 
    \env \turns \rulet \leadsto E
\end{equation*} 

  Based on Lemma~\ref{lemma:resolution} we then know
\begin{equation*} 
    |\env| \turns E : |\rulet|
\end{equation*} 

  Hence, because all variables are unique
\begin{equation*} 
    |\tenv|, |\env| \turns E : |\rulet|
\end{equation*} 

\item[(\texttt{TrRule})\quad$\tenv | \env \turns \ruleabs{\rulet}{e} : \rulet \leadsto \Lambda \vec{\alpha}.\lambda(\vec{x}:|\vec{\rulet}|).E$]\ \\

  Based on (\texttt{TrRule}) and the induction hypothesis, we have
\begin{equation*} 
    |\tenv|,|\env|,\vec{x}:|\vec{\rulet}| \turns E : |\type|
\end{equation*}

  where
\begin{equation*} 
     \rulet = \forall \bar{\alpha}. \bar{\rulet} \To \type
\end{equation*}

  Thus, based on (\texttt{F-Abs}) we have
\begin{equation*} 
    |\tenv|,|\env| \turns \lambda(\vec{x}:|\vec{\rulet}|).E : |\rulet_1| \rightarrow \ldots \rightarrow |\rulet_n| \rightarrow |\type|
\end{equation*}

  or, using the definition of $|\cdot|$
\begin{equation*} 
    |\tenv|,|\env| \turns \lambda(\vec{x}:|\vec{\rulet}|).E : |\bar{\rulet} \To \type|
\end{equation*}

  Moreover, because of (\texttt{TrRule}), we know
\begin{equation*} 
    \vec{\alpha} \cap \mathit{ftv}(\tenv,\env) = \emptyset
\end{equation*}

  and hence
\begin{equation*} 
    \vec{\alpha} \cap \mathit{ftv}(|\tenv|,|\env|) = \emptyset
\end{equation*}

  So, finally, we may conclude from (\texttt{F-TAbs}) that
\begin{equation*} 
    |\tenv|,|\env| \turns \Lambda \vec{\alpha}.\lambda(\vec{x}:|\vec{\rulet}|).E : \forall \vec{\alpha}.|\bar{\rulet} \To \type|
\end{equation*}
 
  and again with $|\cdot|$
\begin{equation*} 
    |\tenv|,|\env| \turns \Lambda \vec{\alpha}.\lambda(\vec{x}:|\vec{\rulet}|).E : |\forall \vec{\alpha}.\bar{\rulet} \To \type|
\end{equation*}
 
\item[(\texttt{TrInst})]\quad$\tenv | \env \turns e[\vec{\type}] : [\vec{\alpha} \mapsto \vec{\type}](\bar{\rulet} \To \type) \leadsto E\,|\vec{\type}|$\ \\
 
  By (\texttt{TrInst}) and the induction hypothesis, it follows that
\begin{equation*} 
    |\tenv|, |\env| \turns E : |\forall \vec{\alpha}.\bar{\rulet} \To \type|
\end{equation*} 

  From which we have by definition of $|\cdot|$
\begin{equation*} 
    |\tenv|, |\env| \turns E : \forall \vec{\alpha}.|\bar{\rulet} \To \type|
\end{equation*} 

  It follows from (\texttt{F-TApp}) that
\begin{equation*} 
    |\tenv|, |\env| \turns E\,|\vec{\type}| : [\vec{\alpha} \mapsto |\vec{\type}|]|\bar{\rulet} \To \type|
\end{equation*} 

  which is easily seen to be equal to 
\begin{equation*} 
    |\tenv|, |\env| \turns E\,|\vec{\type}| : |[\vec{\alpha} \mapsto \vec{\type}](\bar{\rulet} \To \type)|
\end{equation*} 

\item[(\texttt{TrRApp})\quad$\tenv | \env \turns \ruleapp{e}{\overline{e:\rulet}} : \type \leadsto E\, \vec{E}$] \ \\

  From (\texttt{TrApp}) and the induction hypothesis we have:
\begin{equation*}
    |\tenv|, |\env| \turns E : |\bar{\rulet} \To \type|
\end{equation*}

  and
\begin{equation*}
    |\tenv|, |\env| \turns E_i : |\rulet_i|   (\forall i)
\end{equation*}

  Hence, base on the definition of $|\cdot|$ the first of these means 
\begin{equation*}
    |\tenv|, |\env| \turns E : |\rulet_1| \rightarrow \ldots \rightarrow |\rulet_n| \rightarrow |\type|
\end{equation*}

  Hence, based on (\texttt{F-App}) we know
\begin{equation*}
    |\tenv|, |\env| \turns E\,\vec{E} : |\type|
\end{equation*}

\end{description}
\end{proof}
{\centering
\fbox{
\begin{minipage}{0.95\columnwidth}
\begin{lemma}\label{lemma:resolution}
  If 
\begin{equation*}
    \env \turns \rulet \leadsto E
\end{equation*}
  then
\begin{equation*}
    |\env| \turns E : |\rulet|
\end{equation*}
\end{lemma}
\end{minipage}
}}

\begin{proof}

  By induction on the derivation.

  From (\texttt{TrRes}) we have
\begin{equation*}
    \env \turns \rulet \leadsto \Lambda\vec{\alpha}.\lambda(\vec{x} : |\vec{\rulet}|).(E\,\vec{E})
\end{equation*}

  where
\begin{equation*}
    \rulet = \forall \vec{\alpha}.\bar{\rulet} \To \type
\end{equation*}

  Also from (\texttt{TrRes}) and the induction hypothesis, we have
\begin{equation*}
    |\env| \turns E_i : |\rulet_i| \quad(\rulet_i \in \bar{\rulet}' - \bar{\rulet})  
\end{equation*}

  Also from (\texttt{TrRes}) and Lemma~\ref{lemma:lookup1}, we have
\begin{equation*}
   |\env| \turns E : |\vec{\rulet}' \To \type| 
\end{equation*}

  Assembling these parts using (\texttt{F-App}), (\texttt{F-Abs}) and (\texttt{F-TAbs}) we come to
\begin{equation*}
   |\env| \turns \Lambda\vec{\alpha}.\lambda(\vec{x}:|\vec{\rulet}|).(E\,\vec{E})
\end{equation*}

\end{proof}
{\centering
\fbox{
\begin{minipage}{0.95\columnwidth}
\begin{lemma}\label{lemma:lookup1}
  If 
\begin{equation*}
    \lookup{\env}{\type} = \rulesetvar \To \type : E
\end{equation*}
  then
\begin{equation*}
    |\env| \turns E : |\rulesetvar \To \type|
\end{equation*}
\end{lemma}
\end{minipage}
}}

\begin{proof}
  This follows trivially from Lemma~\ref{lemma:lookup2}.
\end{proof}

{\centering
\fbox{
\begin{minipage}{0.95\columnwidth}
\begin{lemma}\label{lemma:lookup2}
  If 
\begin{equation*}
    \lookup{\overline{\rulet:x}}{\type} = \rulesetvar \To \type : E
\end{equation*}
  then
\begin{equation*}
    |\overline{\rulet:x}| \turns E : |\rulesetvar \To \type|
\end{equation*}
\end{lemma}
\end{minipage}
}}

\begin{proof}

  From the definintion of lookup we know that iff
\begin{equation*}
    \lookup{\overline{\rulet:x}}{\type} = \theta \rulesetvar' \To \type : x \, |\vec{\type}| 
\end{equation*}

  then 
\begin{equation*}
    (\rulet : x) \in \overline{\rulet:x}
\end{equation*}

  Hence, it trivially follows that
\begin{equation*}
    (x : |\rulet|) \in |\overline{\rulet:x}|
\end{equation*}

  Hence, from (\texttt{F-Var}) we have that
\begin{equation*}
    |\overline{\rulet:x}| \turns x : |\rulet|
\end{equation*}

  Following the definition of $|\cdot|$ we also know
\begin{equation*}
    |\overline{\rulet:x}| \turns x : \forall \vec{\alpha}.|\rulesetvar' \To \type'|
\end{equation*}

  So
\begin{equation*}
    |\overline{\rulet:x}| \turns x \, |ts| : |\theta (\rulesetvar' \To \type')|
\end{equation*}
\end{proof}